\documentclass[11pt]{article}

\makeatletter
\renewcommand{\thefootnote}{}

\usepackage{epsfig,amsmath,graphicx,amssymb,overpic}

\usepackage{setspace}
\usepackage{graphicx}
\usepackage{dcolumn}
\usepackage{bm}
\usepackage{graphicx}
\usepackage{subfigure}
\usepackage{epsfig,amsmath,graphicx,amssymb,overpic,cite}

\usepackage{graphicx}
\usepackage{dcolumn}
\usepackage{bm}
\usepackage{graphicx}
\usepackage{subfigure}
\usepackage{amsthm}

\usepackage{pdfsync}
\usepackage{cite}
\usepackage{amscd}
\usepackage{amsmath}
\usepackage{latexsym}
\usepackage{amsfonts}
\usepackage{amssymb}
\usepackage{amsthm}
\usepackage{graphicx}
\usepackage{indentfirst}
\usepackage{enumerate}
\usepackage{amstext}
\usepackage[dvipdfm,
            pdfstartview=FitH,
            CJKbookmarks=true,
            bookmarksnumbered=true,
            bookmarksopen=true,
            colorlinks=true, 
            pdfborder=001,   
            citecolor=magenta,
            linkcolor=blue,
            ]{hyperref}

 \allowdisplaybreaks[4]

\newtheorem{remark}{Remark}[section]
\newtheorem{theorem}{Theorem}[section]

\usepackage{tikz}
\usetikzlibrary{shapes,arrows,positioning}
\usepackage{xcolor} 

\def\be{\begin{equation}}
\def\ee{\end{equation}}
\def\bee{\begin{eqnarray}}
\def\ene{\end{eqnarray}}
\def\bes{\begin{subequations}}
\def\ees{\end{subequations}}

\def\d{\displaystyle}
\def\v{\vspace{0.05in}}

\begin{document}

\baselineskip=13pt
\renewcommand {\thefootnote}{\dag}
\renewcommand {\thefootnote}{\ddag}
\renewcommand {\thefootnote}{ }

\pagestyle{plain}

\begin{center}
\baselineskip=16pt \leftline{} \vspace{-.3in} {\Large \bf Long-time asymptotics for the $N_{\infty}$-soliton solution to the KdV equation with two types of generalized reflection coefficients
} \\[0.2in]
\end{center}

\begin{center} \small
{\bf Guoqiang Zhang$^{a}$, Zhenya Yan}$^{a,b,*}$\footnote{$^{*}${\it Email address}: zyyan@mmrc.iss.ac.cn (Corresponding author)}  \\[0.1in]
{\it$^a${\footnotesize KLMM,  Academy of Mathematics and Systems Science,  Chinese Academy of Sciences, Beijing 100190, China}}\\
{\it $^b$\footnotesize School of Mathematical Sciences, University of Chinese Academy of Sciences, Beijing 100049, China} \\[0.18in]
\end{center}


\noindent {\bf Abstract:}\, {\small
We systematically investigate the long-time asymptotics for the $N_{\infty}$-soliton solution to the KdV equation in the different regions with the aid of  the Riemann-Hilbert (RH) problems with two types of generalized reflection coefficients on the interval $\left[\eta_1, \eta_2\right]\in \mathbb{R}^+$:
\begin{itemize}
\item $r_0(\lambda,\eta_0; \beta_0, \beta_1,\beta_2)=\left(\lambda-\eta_1\right)^{\beta_1}\left(\eta_2-\lambda\right)^{\beta_2}\left|\lambda-\eta_0\right|^{\beta_0}\gamma\left(\lambda\right)$,
\item $r_c(\lambda,\eta_0; \beta_1,\beta_2)=\left(\lambda-\eta_1\right)^{\beta_1}\left(\eta_2-\lambda\right)^{\beta_2}\chi_c\left(\lambda, \eta_0\right)\gamma
    \left(\lambda\right)$,
\end{itemize}
where the singularity $\eta_0\in (\eta_1, \eta_2)$ and $\beta_j>-1$ for $j=0, 1, 2$, the function $\gamma: \left[\eta_1, \eta_2\right] \to\mathbb{R}^+$ is continuous and positive  on $\left[\eta_1, \eta_2\right]$, with an  analytic extension to a neighborhood of this interval, and the step-like function $\chi_c$ is defined as $\chi_c\left(\lambda,\eta_0\right)=1$ for $\lambda\in\left[\eta_1, \eta_0\right)$ and $\chi_c\left(\lambda,\eta_0\right)=c^2$ for $\lambda\in\left(\eta_0, \eta_2\right]$ with $c>0, \, c\ne1$. A critical step in the analysis of RH problems via  the Deift-Zhou steepest descent technique is how to  construct local parametrices around the endpoints $\eta_j$'s and the singularity $\eta_0$.
Specifically, the modified Bessel functions of indexes $\beta_j$'s are utilized for the endpoints $\eta_j$'s, and
the modified Bessel functions of index $\left(\beta_0\pm 1\right)\left/\right.2$ and confluent hypergeometric functions are employed around the singularity $\eta_0$ if the reflection coefficients are $r_0$ and $r_c$, respectively. This comprehensive study  extends the understanding of generalized reflection coefficients and  provides valuable insights into the asymptotics of soliton gases.}



\vspace{0.1in} \noindent {\small {\bf Keywords}\, KdV equation; Riemann-Hilbert problem; Generalized reflection coefficients; $N$-soliton; Deift-Zhou steepest descent method; Long-time asympotics; Confluent hypergeometric function; Modified Bessel funtion

\vspace{0.1in} \noindent {\bf Mathematics Subject Classification}\, 35Q51, 35Q15, 37K40, 37K10, 37K15





\begin{spacing}{1}
\tableofcontents
\end{spacing}


\baselineskip=14pt

\section{Introduction}
\label{}

Solitons~\cite{soliton}, as a special type of nonlinear wave phenomena, play an significant role in many fields of nonlinear sciences, such as fluid mechanics, nonlinear optics, Bose-Einstein condensates, plasma physics, and etc.~\cite{s1,s2,s4,s5,s7,s8,s9,s10,s11,s-12,s-13}. One of the key issues about soliton phenomena is to explore interaction behaviors among multi-soliton solutions of nonlinear wave equations (particular nonlinear integrable systems)~\cite{si-1,si-2,si-3}.  In 1971, Zakharov~\cite{1} first introduce the concept of a {\it soliton gas} (i.e., the limit behavior of $N$-soliton solution as $N\to \infty$) to characterize an infinite ensemble of weakly interacting solitons within the framework of the integrable Korteweg-de Vries (KdV) equation~\cite{28}
\begin{gather}\label{KdV}
u_t - 6uu_x + u_{xxx} = 0, \quad u=u(x,t),\quad (x,t)\in \mathbb{R}\times \mathbb{R}^+,
\end{gather}
where a kinetic equation was formulated to model the evolution of spectral distribution functions associated with this ensemble~\cite{miura}. Building upon Zakharov's conceptualization of a rarefied gas, the kinetic equation was later generalized to a dense KdV soliton gas \cite{2}. This generalization involved developing spectral theory via the thermodynamic limit of finite-gap solutions, which was subsequently applied to a soliton gas \cite{3} and a breather gas \cite{4} for the focusing nonlinear Schr\"odinger (NLS) equation, and a soliton gas in bidirectional dispersive hydrodynamics \cite{5} for the defocusing NLS equation. Beyond the spectral theory, significant attention has been paid to the mathematical properties of soliton gases~\cite{el2021}, including integrable reductions \cite{6} and hydrodynamic reductions \cite{7} for their kinetic equations, minimal energy solutions \cite{8}, and the classical integrability of hydrodynamics \cite{9}, among others. In the realm of numerical simulations, it has been suggested that the nonlinear phase of spontaneous modulational instability \cite{10} and the emergence of rogue waves \cite{11,12} can be fundamentally attributed to the dynamics of soliton gases. Recently, some physical experiments were present to realize a bidirectional soliton gas in a 34-m-long wave flume in a shallow water regime, in which the corresponding integrable model is Kaup-Boussinesq system~\cite{gas-exp}, and optical soliton gas~\cite{gas-exp2}.

The Riemann-Hilbert (RH) problems associated to the inverse scattering transform (IST)~\cite{28} play an important role in the study of nonlinear integrable systems. It can be used to not only rigorously solve solitons of integrable systems~\cite{28,s-12,z1,z2,con1,con2}, but also to analyze long-time asymptotics of solutions of some integrable systems with some types of initial data, such as the modified KdV (mKdV) equation, NLS equation, derivative NLS equation, Camassa-Holm (CH) equation, modified CH equation, Fokas-Lenells equation, sine-Gordon equations, and short-pulse equation (see, e.g., Refs.~\cite{16,d1,d2,d3,d4,d5,d6,d7,b1,b2,b3,b4,b5,McL1,McL2} and references therein) via the Deift-Zhou steepest descent technique~\cite{16}, or the $\bar\partial$-method~\cite{McL1,McL2}. Moreover, Fokas~\cite{f1} presented a unified method to use the RH problem to express the solutions of linear and nonlinear integrable systems with initial-boundary value conditions~\cite{f2,f3,f4}.
In recent years, the asymptotic analysis of soliton gases has garnered  significant attention and rapidly increasing interest, particularly through the use of RH problems. The soliton gas for the KdV equation \eqref{KdV}, referred to as the primitive potential, is described using the dressing method \cite{13} and a RH problem \cite{14}, which involves two reflection coefficients in the jump conditions. When only one reflection coefficient is considered, a thorough investigation \cite{15} of the asymptotic behavior is conducted using the Deift-Zhou steepest descent technique~\cite{16} and further developed in \cite{17,18,19}. More recently, the method for deriving the asymptotics of a soliton gas has been applied and extended to the mKdV equation \cite{20} to study the interaction between a soliton gas and a large soliton, as well as to describe various wave properties. These properties include the local phase shift of the soliton gas, the location of the soliton peak, and notably, the average velocity of the soliton peak modeled by a kinetic equation. Beyond the discrete spectra confined to  a segment on the real axis $\mathbb{R}$ or the purely imaginary axis $\mathrm{i}\mathbb{R}$, soliton gases derived from the $N$-soliton solution in the limit as $N\to\infty$ have been extended to the focusing NLS equation with the discrete spectra being located in bounded domains, revealing the  surprising  soliton shielding effect\cite{21}.

It should be noted that the reflection coefficient $r(\lambda)$ in \cite{15} is assumed to:  i) be a continuous and strictly positive function for $\lambda\in\left[\eta_1, \eta_2\right]$;  ii) be  analytically extended to a neighborhood near the interval $\left[\eta_1, \eta_2\right]$;  and iii)  have the symmetry $r(\lambda)=r(-\lambda)$ for $\lambda\in\left[-\eta_2, -\eta_1\right]$.
These assumptions ensure that the solution $Y$ of the initial RH problem exhibits the local behavior characterized by logarithmic singularities at the endpoints
\begin{gather}\label{log-singularity}
Y=
\begin{cases}
\mathcal{O}
\begin{pmatrix}
\log\left|\lambda-\eta_j\right| &1
\end{pmatrix}
, & \mathrm{as} \,\,\,\lambda\to\eta_j,\\[0.5em]
\mathcal{O}
\begin{pmatrix}
1 & \log\left|\lambda+\eta_j\right|
\end{pmatrix}
, & \mathrm{as} \,\,\,\lambda\to-\eta_j,
\end{cases}
\end{gather}
with the \( \mathcal{O} \)-term interpreted element-wise.
Consequently,  the local parametrices near the endpoints are expressed in terms of modified Bessel functions of index $0$.
Additionally, a special case was considered in \cite{20} where $r(\lambda)\left|\lambda\mp\eta_j\right|^{\pm1/2}=\mathcal{O}\left(1\right)$ at $\lambda=\pm\eta_j$.
In this case, local parametrices at the endpoints $\lambda=\eta_j$ are unnecessary, as the outer parametrix sufficiently matches local behaviors and can serve as the global parametrix.
The outer parametrix was initially developed  to study the long-time asymptotics of the KdV equation with step-like initial values  \cite{22}.
Solving the matrix RH problem with the same jump matrices, the matrix outer parametrix was first introduced in \cite{23} for the  asymptotic analysis of orthogonal polynomials   related to Hermitian matrix model. It was subsequently applied to long-time asymptotics for the NLS shock problem \cite{24}, the mKdV equation with step-like initial data \cite{25},  and the NLS equation with nonzero boundary conditions \cite{26,27}.

Inspired by the significant contributions about soliton gases with RH problems in \cite{15,20} and modified Jacobi weights of orthogonal polynomials~\cite{61,57,68}, in this paper, we would like to introduce two types of generalized reflection  coefficients in the KdV solitons, and propose the corresponding long-time asymptotic behaviors of soliton gases. The first type of generalized reflection coefficient is
\begin{gather}\label{r0}
r_0(\lambda,\eta_0; \beta_0,\beta_1,\beta_2)=\left(\lambda-\eta_1\right)^{\beta_1}\left(\eta_2-\lambda\right)^{\beta_2}\left|\lambda-\eta_0\right|^{\beta_0}\gamma\left(\lambda\right),
\end{gather}
where the singularity $\eta_0\in (\eta_1, \eta_2)$ and $\beta_j>-1$ for $j=0, 1, 2$, and $\gamma(\lambda)$, as the same setting of the reflection coefficient  in \cite{15}, is a continuous and strictly positive function in $\lambda$ for $\lambda\in\left[\eta_1, \eta_2\right]$ and is assumed to be analytic in a neighborhood near the interval $\left[\eta_1, \eta_2\right]$.
Unlike the original reflection coefficient, which is strictly positive at the endpoints, the generalized reflection coefficient $r_0(\lambda,\eta_0; \beta_0,\beta_1,\beta_2)$ defined by Eq.~(\ref{r0}) has zeros and singularities at $\eta_1$ and $\eta_2$.
Additionally, $r_0(\lambda,\eta_0; \beta_0,\beta_1,\beta_2)$ exhibits the zero and singularity at $\eta_0$ within the interval $\left(\eta_1, \eta_2\right)$, making it significantly different from the known reflection coefficient~\cite{15}.
The absolute value is used to ensure positivity in the subintervals  $\left(\eta_1, \eta_0\right)$ and $\left(\eta_0, \eta_2\right)$. Conversely, the second generalized reflection coefficient considered in this paper is defined as
\begin{gather}\label{rc}
r_c(\lambda,\eta_0; \beta_1,\beta_2)=\left(\lambda-\eta_1\right)^{\beta_1}\left(\eta_2-\lambda\right)^{\beta_2}\chi_c\left(\lambda,\eta_0\right)\gamma\left(\lambda\right).
\end{gather}
While preserving the same behaviors at endpoints as $r_0(\lambda,\eta_0; \beta_0,\beta_1,\beta_2)$,  $r_c(\lambda,\eta_0; \beta_1,\beta_2)$ also has a singularity at $\eta_0\in (\eta_1, \eta_2)$.
Unlike $r_0$, the second generalized reflection coefficient $r_c$ exhibits a jump discontinuity at $\eta_0$, characterized by $\chi_c$.
Specifically, $\chi_c$ is a step-like function defined as follows: $\chi_c\left(\lambda,\eta_0\right)=1$ for $\lambda\in\left[\eta_1, \eta_0\right)$ and $\chi_c(\lambda,\eta_0)=c^2$ for $\lambda\in\left(\eta_0, \eta_2\right]$, where $c$ is a positive constant with $c\ne1$.
The powers $\beta_j$ are constrained by $\beta_1, \beta_2, \beta_0>-1$, with the rationale to be explained later.

\begin{remark} Notice that the introduced new singularity point $\eta_0\in (\eta_1, \eta_2)$ may be understood by the following physical meaning: For the considered soliton gas in some region (interval): $\lambda\in\left[\eta_1, \eta_2\right]$, if we insert an additional singularity point $\eta_0$ in the interval such that the soliton gas is separated into two parts, then a natural and interesting problem is how to describe the new status of the soliton gas ? The original idea introducing the singularity point arises from the modified Jacobi weights of orthogonal polynomials~\cite{61,57,68}.
\end{remark}

The KdV  equation \eqref{KdV} is a completely integrable physical model, and its integrability can be confirmed by the inverse scattering transform with the Lax pair~\cite{28,29,30}
\bee
\left\{\begin{array}{l}
\left(\partial_x^2-u\right)\psi=\lambda^2\psi, \v\\
\psi_t=\left(2u-4\lambda^2\right)\psi_x-u_x\psi,
\end{array}\right.
\ene
where $\psi=\psi(x,t;\lambda)$ is the eigenfunction, and $\lambda\in\mathbb{C}$ is an iso-spectral parameter.
The $N$-soliton solution of the KdV equation \eqref{KdV} is associated with a  RH problem; for an illustration,  refer to the work of Grunert and Teschl \cite{31}.
By slightly modifying and applying an interpolation method that transforms poles into jump conditions, as suggested by Deift {\it et al}~\cite{32} in the study of the Toda rarefaction problem, a special sequence of  $N$-soliton solutions $u_N$ can be reconstructed by
\begin{gather}
\begin{pmatrix}
1&1
\end{pmatrix}
u_N=2\frac{\partial}{\partial x}\left(\lim_{\lambda\to\infty}\lambda\left(M^N(\lambda)-
\begin{pmatrix}
1&1
\end{pmatrix}
\right)\right)
\sigma_3,\quad N=N_1+N_2,
\end{gather}
where $\sigma_3$ is the third Pauli matrix expressed, and $M^N(\lambda)$ is a $1\times 2$ vector-valued function with the following properties: $M^N$ is analytic in $\lambda$ for $\lambda\in\mathbb{C}\setminus\left(\Gamma_+\cup\Gamma_-\right)$, normalizes to
$\begin{pmatrix}
1&1
\end{pmatrix}$
as $\lambda\to\infty$,
and its continuous boundary values  are related by
\begin{gather}\label{MN}
M^N_+=M^N_-
\begin{cases}
\mathcal{L}^{t\theta}\left[\d\sum_{j=1}^{N_1}\frac{\left(\eta_0-\eta_1\right)\,r\left(\lambda_{1, j}\right)}{2N_1\pi\,\left(\lambda-\lambda_{1, j}\right)}
+\sum_{j=1}^{N_2}\frac{\left(\eta_2-\eta_0\right)\,r\left(\lambda_{2, j}\right)}{2N_2\pi\,\left(\lambda-\lambda_{2, j}\right)}
\right], &\mathrm{for}\,\,\, \lambda\in\Gamma_+,\\[2em]
\mathcal{U}^{t\theta}\left[\d -\sum_{j=1}^{N_1}\frac{\left(\eta_0-\eta_1\right)\,r\left(-\lambda_{1, j}\right)}{2N_1\pi\,\left(\lambda+\lambda_{1, j}\right)}
-\sum_{j=1}^{N_2}\frac{\left(\eta_2-\eta_0\right)\,r\left(-\lambda_{2, j}\right)}{2N_2\pi\,\left(\lambda+\lambda_{2, j}\right)}\right],  &\mathrm{for}\,\,\, \lambda\in\Gamma_-,
\end{cases}
\end{gather}
where $\Gamma_\pm$ are two non-intersecting simple closed curves surrounding $\left[\eta_1, \eta_2\right]$ and $\left[-\eta_2, -\eta_1\right]$, respectively, oriented counter-clockwise. The notations in jump matrices are introduced in \eqref{LU}, and  the phase $\theta$  is given by
\begin{gather}
\theta=4\lambda\left(\xi-\lambda^2\right), \quad \xi=\frac{x}{4t}\,\, ({\rm wave\,\, velocity}).
\end{gather}
Discrete spectra $\lambda_{1, j}$ are located in $\left(\eta_1, \eta_0\right)$ with equal spacing, defined as $\lambda_{1, j}=\eta_1+j(\eta_0-\eta_1)/(N_1+1)$ for $j=1, 2, \cdots N_1$.
Similarly, $\lambda_{2, j}$ are in $\left(\eta_0, \eta_2\right)$, defined as $\lambda_{2, j}=\eta_0+j(\eta_2-\eta_0)/(N_2+1)$ for $j=1, 2, \cdots N_2$.
The discrete spectra are located within the interval $\left(\eta_1, \eta_2\right)$, allowing the reflection coefficient $r$ to have singularities of order $\beta_j$ at the endpoints.
Unlike in \cite{15}, the absence of $\eta_0$ in the discrete spectra distribution permits more local behaviors at $\lambda=\eta_0$, such as the singularity of order $\beta_0$ in $r_0$ and the jump discontinuity in $r_c$.
It is important to note that two types of reflection coefficients are considered, namely, $r=r_0$ and $r=r_c$, with the values of $r$ at $\left[-\eta_2, -\eta_1\right]$ determined by the symmetry $r(\lambda)=r(-\lambda)$.

The limit technique is a powerful and effective tool for characterizing novel nonlinear wave solutions  that cannot be derived by direct methods in the context of  integrable systems.
For example, the infinite-order rogue wave can be captured \cite{33} by taking the limit of a sequence of RH problems associated with $N$th-order rogue waves in the context of the robust inverse scattering transform \cite{34}.
This method succeeds where the Darboux transformation \cite{35,36,37}, Hirota bilinear method \cite{38,39}, and the inverse scattering transform \cite{40,41} fail to explicitly derive the solution.
The same limit technique is applied to capture an infinite-order soliton solution by means of a RH problem for the focusing NLS equation \cite{42}.
In the above limit procedure, an appropriate rescaling transform plays a crucial role in the convergence of the sequences of  jump matrices.
Additionally, it makes the study of large-$N$ asymptotics of $N$-soliton solutions with the initial value $N\mathrm{sech}(x)$ equivalent to a semiclassical limit problem \cite{43}.
In contrast to appropriate rescaling transforms, the key step in \cite{15} involves taking appropriate norming constants, leading the limit to converge to a  definite Riemann integral.
In this paper, the norming constants are taken as the discretization of generalized reflection coefficients $r$, and  the limit for obtaining a soliton gas  converges not only to definite integrals but also to  improper integrals.
Alternatively, the limit process in \cite{15} can be interpreted as  substituting the reflection coefficient with its semiclassical approximation, as discussed in \cite{44,45}.

By taking $N_1\to\infty$ and $N_2\to\infty$, the RH problem associated with a soliton gas is obtained.
The $1\times 2$ vector-valued solution, denoted as $M^\infty$, satisfies the following properties:
$M^\infty$ is analytic for $\lambda\in\mathbb{C}\setminus\left(\Gamma_+\cup\Gamma_-\right)$; it normalizes to $M^\infty\to
\begin{pmatrix}
1&1
\end{pmatrix}$,
as $\lambda\to\infty$;
and its continuous boundary values are related by
\begin{gather}\label{Minfty}
M^\infty_+=M^\infty_-
\begin{cases}
\mathcal{L}^{t\theta}\left[-\mathrm{i}\left(\mathcal{P}_1+\mathcal{P}_2\right)
\right], &\mathrm{for}\,\,\, \lambda\in\Gamma_+,\\[0.5em]
\mathcal{U}^{t\theta}\left[\mathrm{i}\left(\mathcal{P}_{-1}+\mathcal{P}_{-2}\right)\right],  &\mathrm{for}\,\,\, \lambda\in\Gamma_-,
\end{cases}
\end{gather}
where
\bee \mathcal{P}_1=\int_{\eta_1}^{\eta_0}\frac{r(s)}{s-\lambda}\mathrm{d}s, \quad \mathcal{P}_2=\int_{\eta_0}^{\eta_2}\frac{r(s)}{s-\lambda}\mathrm{d}s, \quad \mathcal{P}_{-1}=\int_{-\eta_0}^{-\eta_1}\frac{r(s)}{s-\lambda}\mathrm{d}s, \quad \mathcal{P}_{-2}=\int_{-\eta_2}^{-\eta_0}\frac{r(s)}{s-\lambda}\mathrm{d}s.
\ene
Similar to the use of Zhou's vanishing lemma \cite{46} for the RH problem 7 in \cite{20},  the unique solution for $M^\infty$ can also be obtained.
As stated  in \cite{14,15}, the soliton gas $u$ can be recovered from $M^\infty$ by
\begin{gather}
\begin{pmatrix}
1&1
\end{pmatrix}
u=2\frac{\partial}{\partial x}\left(\lim_{\lambda\to\infty}\lambda\left(M^\infty(\lambda)-
\begin{pmatrix}
1&1
\end{pmatrix}
\right)\right)
\sigma_3.
\end{gather}
In this limit procedure, it is crucial that the jump matrices from \eqref{MN} to \eqref{Minfty} are valid if the exponential coefficients $\beta_1$, $\beta_2$, and $\beta_0$ satisfy $\beta_1, \beta_2, \beta_0>-1$.
To clarify this, two cases are considered below.
First, if $\beta_1, \beta_2, \beta_0\ge 0$, this validity is evident via the definition of  the definite Riemann integral.
Secondly, if $\beta_1\in\left(-1, 0\right)$,  $\beta_2\in\left(-1, 0\right)$,  or $\beta_0\in\left(-1, 0\right)$, improper integrals of the second kind arise due to singularities, and hence the convergence of the Riemann sum cannot be inferred from the definition of definite integrals.
Nonetheless, it can still be proven that the special Riemann sum converges to improper integrals.
Indeed, the proof  only involves calculus and can be achieved with the help of monotonicity and  uniform continuity.

\section{Statement  of main results}

Following the approach in \cite{15},  several transforms $Y\mapsto T\mapsto S\mapsto E$ are applied to ensure that $E$ normalizes to $\begin{pmatrix}1&1 \end{pmatrix}$ at infinity and that its jump matrices decay exponentially and uniformly  to the identity matrix. This paper presents the long-time asymptotics of KdV soliton gases with two different types of generalized reflection coefficients $r_0$ and $r_c$ in these different regions $\left(-\infty, \xi_\mathrm{crit}\right)$, $\left(\xi_\mathrm{crit}, \xi_0\right)$, $\left(\xi_0, \eta_2^2\right)$, and $\left(\eta_2^2, +\infty\right)$. The values of $\xi_\mathrm{crit}$ and $\xi_0$ are
uniquely determined by
\begin{gather}
\xi_\mathrm{crit}=\eta_2^2\,W\left(\frac{\eta_1}{\eta_2}\right), \quad \xi_0=\eta_2^2\,W\left(\frac{\eta_0}{\eta_2}\right),
\end{gather}
 where $W: m\mapsto W(m)$ is a function defined by
\begin{gather}
W(m)=\frac{m^2+1}{2}+\frac{m^2\left(m^2-1\right)}{m^2-1+eE(m)/eK(m)},
\end{gather}
with $eE$ and $eK$ being the complete elliptic integrals of the first and second kind, respectively,
\begin{gather}\label{elliptic12}
eK(\lambda)=\int_0^{\pi/2}\frac{\mathrm{d}y}{\sqrt{1-\lambda^2\sin^2 y}}, \qquad  eE(\lambda)=\int_0^{\pi/2}\sqrt{1-\lambda^2\sin^2 y}\,\mathrm{d}y.
\end{gather}
The function $W$ is related to an appropriate so-called $g$-function in the conjugation action for the purpose of exponential decay on lenses, and hence, the Airy parametrix fits the local behavior around $\lambda=\alpha$.
The value of $\alpha$ is uniquely determined by the Whitham evolution equation \cite{47}
\begin{gather}\label{Whitham}
\xi=\eta_2^2\, W\left(\frac{\alpha}{\eta_2}\right),
\end{gather}
which was earlier used by Gurevich and Pitaevskii \cite{48} to characterize a dispersive shock wave for the KdV equation.

\v In the cases of $r=r_0$ and  $r=r_c$, the main results of this paper are summarized as follows.

\v
\begin{theorem} (Long-time asymptotic behaviors of KdV soliton gases)

\begin{enumerate}[0.]

\item[\textbullet] For $\xi\in\left(\xi_0, \eta_2^2\right)$ with  $\beta_0,\, \beta_1\ge 0, \, \beta_2>-1$,
and for  $\xi\in\left(\xi_{\mathrm{crit}}, \xi_0\right)$ with $\beta_0,\, \beta_2>-1,\, \beta_1\ge 0$,
the KdV soliton gas, $u(x, t)$, has  the following long-time asymptotic behavior
\begin{gather}\label{asymptotic-middle}
u(x,t)=u_{0,c}\left(x,t;\alpha\right)+\mathcal{O}\left(\frac{1}{t}\right),
\quad
\mathrm{as}\,\,\,t\to+\infty,
\end{gather}
where the leading term is
\begin{gather}
u_{0,c}\left(x,t;\alpha\right)=-2\alpha^2\mathrm{cn}^2\!\!\left(\eta_2\left[x-2\left(\eta_2^2+\alpha^2\right)t
 +\phi_{0,c}\left(\alpha\right)\right]
+eK\left(\frac{\alpha}{\eta_2}\right)\,\bigg|\ \frac{\alpha}{\eta_2}\right)+\alpha^2-\eta_2^2,
\end{gather}
with $\mathrm{cn}(\cdot|
\cdot)$ being the Jacobi elliptic function, and the phase $\phi$ expressed by
\begin{gather}
\phi_{0,c}(\alpha)=-\frac{1}{\pi}\int_\alpha^{\eta_2}\frac{\log r_{0,c}(s)}{\sqrt{\left(s^2-\alpha^2\right)\left(\eta_2^2-s^2\right)}}\,\mathrm{d}s\in\mathbb{R}.
\end{gather}
Specially, when $\beta_0=0$, the asymptotic formula \eqref{asymptotic-middle} holds for $\xi\in\left(\xi_\mathrm{crit}, \eta_2^2\right)$ with $\beta_1\ge 0, \beta_2>-1$. Note that the different reflection coefficient $r_{0}$ or $r_c$ leads to the change of phase of the double-periodic solution $u_{0,c}$.

\item[\textbullet] For the region $\xi\in\left(-\infty, \xi_\mathrm{crit}\right)$  with $\beta_1, \beta_2, \beta_0>-1$, the long-time  asymptotic behavior of the KdV soliton gas $u(x, t)$ is formulated as
\begin{gather}\label{asymptotic-left}
u=u_{0,c}\left(\eta_1\right)+\mathcal{O}\left(\frac{1}{t}\right),
\quad
\mathrm{as}\,\,\,t\to+\infty.
\end{gather}

\item[\textbullet] For $\xi\in\left(\eta_2^2, +\infty\right)$ with $\beta_1, \beta_2, \beta_0\ge 0$, there exists a positive constant $\mu$ such that the KdV soliton gas $u(x, t)$ has
\begin{gather}\label{asymptotic-right}
u=\mathcal{O}\left(\mathrm{e}^{-\mu t}\right),
\quad
\mathrm{as}\,\,\,t\to+\infty.
\end{gather}
\end{enumerate}

\end{theorem}

\begin{remark} At $\beta_1=\beta_2=\beta_0=0$, the generalized reflection coefficient $r_0$ defined by Eq.~\eqref{r0} reduces to the case considered in \cite{15}. In this scenario, the local parametrix around the endpoints is constructed using the modified Bessel function of index 0, and there is no need to construct a local parametrix around the singularity $\eta_0$.
For $\beta_1=0$ or $\beta_2=0$, the local behaviors and the local parametrix around the corresponding endpoint are detailed in \cite{15}.
For $\beta_0=0$, one can open lenses as described in \cite{15}.
Since the main differences from \cite{15} involve local behaviors and the construction of local parametrices, and since \eqref{asymptotic-left} follows directly from the standard small norm argument, it is sufficient, without loss of generality, to prove \eqref{asymptotic-middle} and \eqref{asymptotic-left} for $\beta_1, \beta_2, \beta_0\in\left(-1, 0\right)\cup\left(0, +\infty\right)$.
\end{remark}

\begin{remark}
In this paper, we consider only one singularity $\eta_0$ for the generalized reflection coefficients $r_0$ and $r_c$. In fact,
these reflection coefficients can also be extended to accommodate an arbitrary number $n$ of singularities $\{\eta_{0,j}\}_1^n$.
For the first type of generalized reflection coefficient, we consider
\begin{gather}\label{general-0}
r_{0,n}=\left(\lambda-\eta_1\right)^{\beta_1}\left(\eta_2-\lambda\right)^{\beta_2}\left(\prod_{j=1}^n\left|\lambda-\eta_{0, j}\right|^{\beta_{0, j}}\right)\gamma\left(\lambda\right),
\end{gather}
where $\eta_1<\eta_{0, 1}<\eta_{0, 2}<\cdots<\eta_{0, n}<\eta_2$ and $\beta_{0, j}\in(-1, 0)\cup (0, +\infty)$.
For the second type of generalized reflection coefficient, we consider
\begin{gather}\label{general-1}
r_{c,n}=\left(\lambda-\eta_1\right)^{\beta_1}\left(\eta_2-\lambda\right)^{\beta_2}\left(\prod_{j=1}^n\chi_j\left(\lambda\right)\right)\gamma\left(\lambda\right),
\end{gather}
where $\chi_j$ is a step-like function defined as $\chi_j\left(\lambda\right)=1$ for $\lambda\in\left[\eta_{0, j-1}, \eta_{0, j}\right)$ and $\chi_j\left(\lambda\right)=c_j^2$ for $\lambda\in\left(\eta_{0, j}, \eta_{0, j+1}\right]$, with $c_j\ne 0$, and $\eta_1=\eta_{0, 0}<\eta_{0, 1}<\eta_{0, 2}<\cdots<\eta_{0, n}<\eta_{0, n+1}=\eta_2$.
The long-time asymptotics of soliton gases characterized by these more general forms \eqref{general-0} and \eqref{general-1} can be obtained by following the steps outlined in this paper, with the addition of local parametrices near $\eta_{0, j}$.
The local parametrix near $\eta_{0, j}$ is constructed using modified Bessel functions with index $(\beta_{0, j}\pm 1)/2$ for the first case \eqref{general-0}, and confluent hypergeometric functions for the second case \eqref{general-1}.

\end{remark}

\begin{remark}
Using an analogous procedure to derive the long-time asymptotics, one can present the large-$x$ asymptotics for the initial data $u(x, 0)$ in the case of  $r=r_0$ and $r=r_c$ with $\beta_0, \beta_1, \beta_2>-1$ as follows.
As $x\to+\infty$, there exists a positive constant $\mu_0$ such that
\begin{gather}\label{asymptotic-right-0}
u(x, 0)=\mathcal{O}\left(\mathrm{e}^{-\mu_0 x}\right).
\end{gather}
As $x\to-\infty$, the initial data $u(x, 0)$ has the following asymptotic
\begin{gather}\label{asymptotic-middle-0}
u(x, 0)=-2\eta_1^2\mathrm{cn}^2\left(\eta_2\left(x+\phi_{0,c}\left(\eta_1\right)\right)
+eK\left(\frac{\eta_1}{\eta_2}\right)\,\bigg|\ \frac{\eta_1}{\eta_2}\right)+\eta_1^2-\eta_2^2+\mathcal{O}\left(\frac{1}{\left|x\right|
}\right).
\end{gather}
\end{remark}

\begin{remark}
The main differences between our paper and Ref.~\cite{15} are as follows:

\begin{itemize}
\item[{\rm (1)}] In our paper, for the configuration of the RH problem for $Y$, the local behaviors at the endpoints $\pm\eta_j$  vary with the value of $\beta_j$, and particularly differ from \eqref{log-singularity} when $\beta_j\ne 0$.
The local behaviors must be classified into three categories: $\beta_j=0$, $\beta_j>0$ and $-1<\beta_j<0$.
Moreover, since the singularity $\eta_0$ appears for both $r=r_0$ and $r=r_c$, the jump conditions on the positive axis are formulated over the interval $\left(\eta_1, \eta_0\right)\cup\left(\eta_0, \eta_2\right)$ instead of  $\left(\eta_1, \eta_2\right)$. Consequently, the local behaviors at $\pm\eta_0$ need to be included.

\item[{\rm (2)}] The reflection coefficient can be analytically extended to a neighborhood of $\left[\eta_1, \eta_2\right]$ in \cite{15}. However, the generalized reflection coefficients ($r_0, r_c$) considered in our paper cannot. Furthermore, the singularity at $\eta_0$ affects the configuration of the opening lenses.
For the  contour deformations of the RH problem for $T$, lenses must be opened above and  below the intervals $\left(\eta_1, \eta_0\right)\cup\left(\eta_0, \eta_2\right)$ for both $r=r_0$ and $r=r_c$.

\item[{\rm (3)}] In our paper, the local parametrices near the endpoints $\pm \eta_j$ are constructed using modified Bessel functions of index $\beta_j$; however, in \cite{15}, they involve modified Bessel functions of index $0$.
Beyond the local behaviors near the endpoints, local parametrices are also needed at the singularities $\pm\eta_0$ for both $r=r_0$ and $r=r_c$. For $r=r_0$, they are constructed using modified Bessel functions of index $\left(\beta_0\pm 1\right)\left/\right.2$, and  for $r=r_c$, they are constructed using confluent hypergeometric functions.
\end{itemize}

\end{remark}

\vspace{1em}
\noindent \textbf{Notation}\,\,
This introduction concludes with a discussion of the notational conventions employed.
The subscripts \( + \) and \( - \) denote non-tangential boundary values from the left and right, respectively, along a jump contour in a RH problem.
Throughout this paper, the three Pauli matrices are frequently utilized
\begin{gather}\label{sigma}
\sigma_1 =
\begin{pmatrix}
0 & 1 \\
1 & 0
\end{pmatrix}, \quad
\sigma_2 =
\begin{pmatrix}
0 & -\mathrm{i} \\
\mathrm{i} & 0
\end{pmatrix}, \quad
\sigma_3 =
\begin{pmatrix}
1 & 0 \\
0 & -1
\end{pmatrix}.
\end{gather}
For conciseness in the formulation of jump matrix, one introduces \( \mathcal{L}^{\lambda_1}_{\lambda_2}\left[\lambda_0\right] \) and \( \mathcal{U}^{\lambda_1}_{\lambda_2}\left[\lambda_0\right] \) as
\begin{gather}\label{LU}
\begin{array}{l}
\mathcal{L}^{\lambda_1}_{\lambda_2}\left[\lambda_0\right] = \mathrm{e}^{\lambda_1\sigma_3} \lambda_2^{-\sigma_3} \mathcal{L}\left[\lambda_0\right] \lambda_2^{\sigma_3} \mathrm{e}^{-\lambda_1\sigma_3}, \v\v \\
\mathcal{U}^{\lambda_1}_{\lambda_2}\left[\lambda_0\right] = \mathrm{e}^{\lambda_1\sigma_3} \lambda_2^{-\sigma_3} \mathcal{U}\left[\lambda_0\right] \lambda_2^{\sigma_3} \mathrm{e}^{-\lambda_1\sigma_3},
\end{array}
\end{gather}
where \( \mathcal{L}\left[\lambda_0\right] \) and \( \mathcal{U}\left[\lambda_0\right] \) are lower and upper triangular matrices, respectively, with all diagonal entries being equal to 1
\begin{gather}\label{LU0}
\mathcal{L}[\lambda_0] =
\begin{pmatrix}
1 & 0 \\
\lambda_0 & 1
\end{pmatrix}, \qquad
\mathcal{U}[\lambda_0] =
\begin{pmatrix}
1 & \lambda_0 \\
0 & 1
\end{pmatrix}.
\end{gather}
Two special cases of \eqref{LU} need to be noted.
On one hand,  $\mathcal{L}^{\lambda_1}$ and $\mathcal{U}^{\lambda_1}$ are a special case of equation \eqref{LU} with $\lambda_2=1$.
On the other hand, $\mathcal{L}_{\lambda_2}$ and $\mathcal{U}_{\lambda_2}$ are a special case of equation \eqref{LU} with $\lambda_1=0$.
The matrices \( C \), \( C_0 \), and \( C_1 \) are three constant matrices
 defined as
\begin{gather}
C = \frac{1}{\sqrt{2}}
\begin{pmatrix}
1 & \mathrm{i} \\
\mathrm{i} & 1
\end{pmatrix}, \qquad
C_0 = -\sqrt{2\pi}
\begin{pmatrix}
1 & 0 \\
0 & \mathrm{i}
\end{pmatrix}, \qquad
C_1 = \frac{1}{\sqrt{2}}
\begin{pmatrix}
1 & -1 \\
1 & 1
\end{pmatrix}.
\end{gather}
Let \( B(\lambda_0) \) denote a neighborhood centered at \( \lambda = \lambda_0 \) on the \( \lambda \)-plane.

\section{Long-time asymptotics: the region $\xi_0<\xi<\eta_2^2$}
\label{}

In this section, we would like to present the long-time asymptotics of the KdV soliton gas $u$ for the region $\xi_0 < \xi < \eta_2^2$.
As stated in the Introduction, it is sufficient, without loss of generality, to prove this for $\beta_0>0, \beta_1>0, \beta_2\in(-1, 0)\cup (0, +\infty)$.

Several transforms, $Y \mapsto T \mapsto S \mapsto E$, are applied to ensure that $E$ normalizes to
$\begin{pmatrix}
1&1
\end{pmatrix}$
at infinity and its jump matrices decay exponentially and uniformly to an identity matrix.
These transformations were performed by Deift, Kriecherbauer, McLaughlin, Venakides, and Zhou for the asymptotics of orthogonal polynomials with respect to exponential weights \cite{23,49}, based on the Riemann-Hilbert problem introduced by Fokas, Its, and Kitaev \cite{50,51}.
Subsequently, the Deift-Zhou steepest descent technique for orthogonal polynomials has been applied to various types of orthogonal polynomials, such as those with logarithmic weight \cite{52,53}, Freud weight \cite{54}, Laguerre
polynomials \cite{55,56,57,58}, measures supported on the plane \cite{59}, Jacobi weight \cite{60}, modified Jacobi weight \cite{61,62}, and discontinuous Gaussian weight \cite{63}, among others.
Furthermore, this technique has been used to solve related topics, including the distribution of the length of the longest increasing subsequence of random permutations \cite{17} and the asymptotic behavior of the discrete holomorphic map $Z^a$ \cite{64}.

\subsection{Riemann-Hilbert problem for $Y$}

The Riemann-Hilbert problem for $M^\infty$ reveals that the jump matrices \eqref{Minfty} are composed of Cauchy integrals. To simplify the problem, one can deform the jump contours and define a $1\times 2$ vector-valued function $Y$ as follows:
\begin{gather}
Y=\begin{cases}
M^\infty \mathcal{L}^{t\theta}\left[-\mathrm{i}\left(\mathcal{P}_1+\mathcal{P}_2\right)
\right], &\mathrm{for}\,\,\, \lambda\,\,\, \text{interior to} \,\,\,\Gamma_+,  \\[0.5em]
M^\infty \mathcal{U}^{t\theta}\left[\mathrm{i}\left(\mathcal{P}_{-1}+\mathcal{P}_{-2}\right)
\right], &\mathrm{for}\,\,\, \lambda \,\,\, \text{interior to} \,\,\, \Gamma_-, \\[0.5em]
M^\infty, & \mathrm{for}\,\,\,\lambda\,\,\, \text{exterior to}\,\,\, \Gamma_+\cup\Gamma_-.
\end{cases}
\end{gather}
It is evident that $Y$ is analytic in $\lambda$ for $\lambda\in\mathbb{C}\setminus\left(\left[-\eta_2, -\eta_1\right]\cup\left[\eta_1, \eta_2\right]\right)$ and normalizes to
$\begin{pmatrix}
1& 1
\end{pmatrix}$
as $\lambda\to\infty$.
For $\lambda\in\left(\eta_1, \eta_0\right)\cup\left(\eta_0, \eta_2\right)\cup\left(-\eta_2, -\eta_0\right)\cup\left(-\eta_0, -\eta_1\right)$,  $Y$ admits continuous boundary values  denoted by $Y_+$ and $Y_-$,  respectively.
Utilizing the Sokhotski-Plemelj formula, these values  are related by the following jump conditions
\begin{gather}
Y_+=Y_-
\begin{cases}
\mathcal{L}^{t\theta}\left[-\mathrm{i}r\right], &\mathrm{for}\,\,\, \lambda\in\left(\eta_1, \eta_0\right)\cup\left(\eta_0, \eta_2\right),\\[0.5em]
\mathcal{U}^{t\theta}\left[\mathrm{i}r\right],  &\mathrm{for}\,\,\, \lambda\in\left(-\eta_2, -\eta_0\right)\cup\left(-\eta_0, -\eta_1\right),
\end{cases}
\end{gather}
where the values of $r$ over the interval $\left(-\eta_2, -\eta_0\right)\cup\left(-\eta_0, -\eta_1\right)$ are determined by the symmetry $r(-\lambda)=r(\lambda)$.
Note that $\pm\eta_j$'s are endpoints, and thus the continuous boundary values $Y_\pm$ cannot be well-defined.
The difference in jump conditions from \cite{15} lies at $\lambda=\pm\eta_0$.
Since $r_0$ and $r_c$ are singular at these points, the limits of $Y$ as $\lambda\to\pm\eta_0$ do not exist.
To ensure a unique solution of $Y$, local behaviors at both the endpoints and $\pm\eta_0$ must be addressed.
Near each endpoint \( \eta_j \) for \( j=1, 2 \), as $\lambda\to\eta_j$, \( Y \) exhibits the following local behavior
\begin{gather}\label{local-Y-eta-1}
Y=
\begin{cases}
\mathcal{O}
\begin{pmatrix}
1 &1
\end{pmatrix}
, & \text{if} \,\,\, \beta_j\in\left(0, +\infty\right),  \\[0.5em]
\mathcal{O}
\begin{pmatrix}
\left|\lambda-\eta_j\right|^{\beta_j} &1
\end{pmatrix}
,  & \text{if} \,\,\, \beta_j\in\left(-1, 0\right),
\end{cases}
\end{gather}
with the \( \mathcal{O} \)-term interpreted element-wise.
As $\lambda\to -\eta_j$, the local behavior of $Y$ near each endpoint \( -\eta_j \) for \( j=1, 2 \) is
\begin{gather}\label{local-Y-eta-2}
Y=
\begin{cases}
\mathcal{O}
\begin{pmatrix}
1 &1
\end{pmatrix}
, & \text{if} \,\,\, \beta_j\in\left(0, +\infty\right),  \\[0.5em]
\mathcal{O}
\begin{pmatrix}
1& \left|\lambda+\eta_j\right|^{\beta_j}
\end{pmatrix}
,  & \text{if} \,\,\, \beta_j\in\left(-1, 0\right).
\end{cases}
\end{gather}
It can be observed that near the endpoints $\pm\eta_j$ with $j=1, 2$, $Y$ has the same local behavior for the two generalized reflection coefficients $r_0$ and $r_c$, while near the $\pm\eta_0$, $Y$ shows different local behaviors for the two cases.
For the first generalized coefficient $r_0$, $Y$ exhibits the following local behaviors:
if $\beta_0>0$, $Y=\mathcal{O}
\begin{pmatrix}
1 &1
\end{pmatrix}$
as $\lambda\to\pm\eta_0$;
if $\beta_0\in\left(-1, 0\right)$,
\begin{gather}\label{local-Y-eta-01}
Y=
\begin{cases}
\mathcal{O}
\begin{pmatrix}
\left|\lambda-\eta_0\right|^{\beta_0} &1
\end{pmatrix}
, & \text{as} \,\,\, \lambda\to\eta_0  \\[0.5em]
\mathcal{O}
\begin{pmatrix}
1& \left|\lambda+\eta_0\right|^{\beta_0}
\end{pmatrix}
,  & \text{as} \,\,\, \lambda\to-\eta_0.
\end{cases}
\end{gather}
For the second generalized coefficient $r_c$, $Y$ has the following behaviors
\begin{gather}\label{local-Y-eta-02}
Y=
\begin{cases}
\mathcal{O}
\begin{pmatrix}
\log\left|\lambda-\eta_0\right| &1
\end{pmatrix}
, & \mathrm{as} \,\,\,\lambda\to\eta_0,\\[0.5em]
\mathcal{O}
\begin{pmatrix}
1 & \log\left|\lambda+\eta_0\right|
\end{pmatrix}
, & \mathrm{as} \,\,\,\lambda\to-\eta_0.
\end{cases}
\end{gather}

\subsection{Riemann-Hilbert problem for $T$}

According to the Deift-Zhou steepest descent method~\cite{16}, the next step is to perform triangular decompositions to facilitate contour deformation
\begin{gather}
\begin{gathered}
\mathcal{L}^{t\theta}\left[-\mathrm{i}r\right]=\mathcal{U}^{t\theta}\left[\mathrm{i}r^{-1}\right]\left(\mathcal{L}^{t\theta}\left[-\mathrm{i}r\right]+\mathcal{U}^{t\theta}\left[-\mathrm{i}r^{-1}\right]-2\,\mathbb{I}_2\right)\mathcal{U}^{t\theta}\left[\mathrm{i}r^{-1}\right],\\[0.5em]
\mathcal{U}^{t\theta}\left[\mathrm{i}r\right]=\mathcal{L}^{t\theta}\left[-\mathrm{i}r^{-1}\right]\left(\mathcal{L}^{t\theta}\left[\mathrm{i}r^{-1}\right]+\mathcal{U}^{t\theta}\left[\mathrm{i}r\right]-2\,\mathbb{I}_2\right)\mathcal{L}^{t\theta}\left[-\mathrm{i}r^{-1}\right].
\end{gathered}
\end{gather}
However, based on the sign of ${\rm Re}\left(\theta\right)$, one can see that exponential decay fails on the corresponding lenses.
Therefore, before we perform the contour deformation, a conjugation operation is needed by introducing an appropriate $g$-function.
The so-called $g$-function method was first developed by Deift,  Venakides, and Zhou in their analysis of the zero-dispersion limit for the KdV equation \cite{18}. One of the most important applications of the $g$-function is to normalize the RH problem \cite{51} of orthogonal polynomials for the order at infinity, constructed via the equilibrium measure \cite{23,49}.  To date, many asymptotic problems have been solved using the  Deift-Zhou steepest descent technique that require the $g$-function.
For instance, the $g$-function was formulated by a conformal mapping from $\mathbb{C}\setminus\left[-1, 1\right]$ onto the exterior of the unit circle for the study of the strong asymptotic of orthogonal polynomials with respect to the modified Jacobi weight \cite{61}. Alternatively, the $g$-function was also conducted from the relevant spectral curve in the study of large-$T$ asymptotics of the infinity-order rogue waves~\cite{33}.

For the KdV soliton gas discussed in this paper, the scalar $g$-function is analytic for $\lambda\in\mathbb{C}\setminus\left[-\eta_2, \eta_2\right]$, and asymptotically behaves as $\mathcal{O}\left(\lambda^{-1}\right)$ as $\lambda\to\infty$. The continuous boundary values $g_\pm$ are related by: $g_++g_- =2\theta$ for $\lambda\in\left(-\eta_2, -\alpha\right)\cup\left(\alpha, \eta_2\right)$;  $g_+-g_-=\Omega$  with  $\Omega=2\mathrm{i}\pi\eta_2(\eta_2^2+\alpha^2-2\xi)/eK\left(\alpha/\eta_2\right)$ for $\lambda\in\left(-\alpha, \alpha\right)$, where $\alpha$ is determined by the Whitham evolution equation \eqref{Whitham}.
It follows from \cite{15,20} that the $g$-function is expressed as
\begin{gather}\label{g-function}
g=\theta+\int_{\eta_2}^\lambda \frac{Q\left(s\right)}{R\left(s\right)}\,\mathrm{d}s,
\end{gather}
where
$Q(s)=12\left(s+\alpha \right)\left(s-\alpha \right)\left(s^2-\left(\eta_2^2-\alpha^2\right)\left/\right.2-\xi\left/\right.3\right)$, and $R(s)$ is a branch of the complex function $\sqrt{\left(s^2-\eta_2^2\right)\left(s^2-\alpha^2\right)}$ such that it is real and positive on $\left(\eta_2, +\infty\right)$ with branch cuts on the contours $\left(\alpha, \eta_2\right)$ and $\left(-\eta_2, -\alpha\right)$.
For more details on the derivation of this type of $g$-function,  refer to  \cite{22,25,65,66}.

In addition to  the $g$-function, a scalar $f$-function is also introduced to achieve constant jump matrices.
The $f$-function is analytic in $\lambda$ for $\lambda\in\mathbb{C}\setminus\left[-\eta_2, \eta_2\right]$ and asymptotically normalizes to 1 as $\lambda\to\infty$.
The continuous boundary values $f_\pm$ are related by: $f_+f_- =r^{-1}$ for $\lambda\in\left(\alpha, \eta_2\right)$;   $f_+f_- =r$ for $\lambda\in\left(-\eta_2, -\alpha\right)$; $f_+f_-^{-1}=\mathrm{e}^\Delta$  with  $\Delta=\int_\alpha^{\eta_2}\left(\log r(s)\right)/R_+\left(s\right)\,\mathrm{d}s\left/\right.\int_0^\alpha 1\left/\right.R(s)\mathrm{d}s\in \mathrm{i}\mathbb{R}$ for $\lambda\in(-\alpha, -\eta_0)\cup(-\eta_0, \eta_0)\cup(\eta_0, \alpha)$.
The $f$-function is obtained by dividing by $R_+$, then taking the logarithm and Plemelj's formula as follows
\begin{gather}\label{f-function}
f=\exp\left\{\frac{R}{2\pi \mathrm{i}}\left(\int_\alpha^{\eta_2}-\frac{\log r(s)}{R_+\left(s\right)}\frac{\mathrm{d}s}{s-\lambda}+\int_{-\eta_2}^{-\alpha}\frac{\log r(s)}{R_+\left(s\right)}\frac{\mathrm{d}s}{s-\lambda}+\int_{-\alpha}^{\alpha}\frac{\Delta}{R(s)}\frac{\mathrm{d}s}{s-\lambda}\right)\right\}.
\end{gather}

With the above preparation, the following conjugation is performed
\begin{gather}\label{Conjugation}
T=Y\mathrm{e}^{tg\sigma_3}f^{\sigma_3},
\end{gather}	
such that  $T$ is a $1\times 2$ vector-valued function  that satisfies a Riemann-Hilbert problem.
$T$ is analytic in $\lambda$ for $\lambda\in\mathbb{C}\setminus\left(\left[-\eta_2, \eta_2\right]\right)$ and normalizes to $
\begin{pmatrix}
1& 1
\end{pmatrix}$
as $\lambda\to\infty$.
For  $\lambda\in\left(-\eta_2, \eta_2\right)\setminus\left\{\pm\eta_1, \pm\eta_0, \pm\alpha\right\}$, $T$ admits continuous boundary values, and they are related by the following jump conditions
\begin{gather}
T_+=T_-
\begin{cases}
\mathcal{U}^{tp_-}_{f_-}\left[\mathrm{i}r^{-1}\right]\left(-\mathrm{i}\sigma_1\right)\mathcal{U}^{tp_+}_{f_+}\left[\mathrm{i}r^{-1}\right], &\mathrm{for}\,\,\, \lambda\in\left(\alpha, \eta_2\right),\\[0.5em]
\mathcal{L}^{tp_-}_{f_-}\left[-\mathrm{i}r^{-1}\right]\left(\mathrm{i}\sigma_1\right)\mathcal{L}^{tp_+}_{f_+}\left[-\mathrm{i}r^{-1}\right], &\mathrm{for}\,\,\, \lambda\in\left(-\eta_2, -\alpha\right),\\[0.5em]
f_-^{-\sigma_3}\mathrm{e}^{tp_-\sigma_3}\mathcal{L}\left[-\mathrm{i}r\right]\mathrm{e}^{-tp_+\sigma_3}f_+^{\sigma_3}, &\mathrm{for}\,\,\, \lambda\in\left(\eta_1, \eta_0\right)\cup\left(\eta_0, \alpha\right),\\[0.5em]
f_-^{-\sigma_3}\mathrm{e}^{tp_-\sigma_3}\mathcal{U}\left[\mathrm{i}r\right]\mathrm{e}^{-tp_+\sigma_3}f_+^{\sigma_3}, &\mathrm{for}\,\,\, \lambda\in\left(-\alpha, -\eta_0\right)\cup\left(-\eta_0, -\eta_1\right),\\[0.5em]
\mathrm{e}^{\left(\Omega t+\Delta\right)\sigma_3}, &\mathrm{for}\,\,\, \lambda\in\left(-\eta_1, \eta_1\right).
\end{cases}
\end{gather}
The local behaviors of $T$ near the endpoints $\pm\eta_1$ and $\pm\eta_0$ are the same as those of $Y$, as long as  $Y$ is replaced by $T$  in Eqs.~\eqref{local-Y-eta-1}-\eqref{local-Y-eta-02}.
However, $T$ exhibits different local behaviors  near the other two endpoints $\pm\eta_2$
\begin{gather}
T=
\begin{cases}
\mathcal{O}
\begin{pmatrix}
 \left|\lambda\mp\eta_2\right|^{\mp\beta_2/2}&\left|\lambda\mp\eta_2\right|^{\pm\beta_2/2}
\end{pmatrix}
, & \mathrm{if} \,\,\,\beta_2\in\left(0, +\infty\right), \v \\
\mathcal{O}
\begin{pmatrix}
 \left|\lambda\mp\eta_2\right|^{\beta_2/2}&\left|\lambda\mp\eta_2\right|^{\beta_2/2}
\end{pmatrix}
, & \mathrm{if} \,\,\,\beta_2\in\left(-1, 0\right).
\end{cases}
\end{gather}
In addition to $\pm\eta_j$ with $j=1, 2, 0$, the local behaviors of $T$ near $\pm\alpha$ should also be presented, as  local parametrices around $\pm\alpha$ need to be constructed. It follows from $g$- and $f$-functions that $T$ has the following local behavior near $\pm\alpha$
\begin{gather}
T=
\mathcal{O}
\begin{pmatrix}
1 &1
\end{pmatrix}
, \quad \mathrm{as} \,\,\,\lambda\to\pm\alpha.
\end{gather}

\subsection{Riemann-Hilbert problem for $S$}

\begin{figure}[!t]
\centering
\includegraphics[scale=0.32]{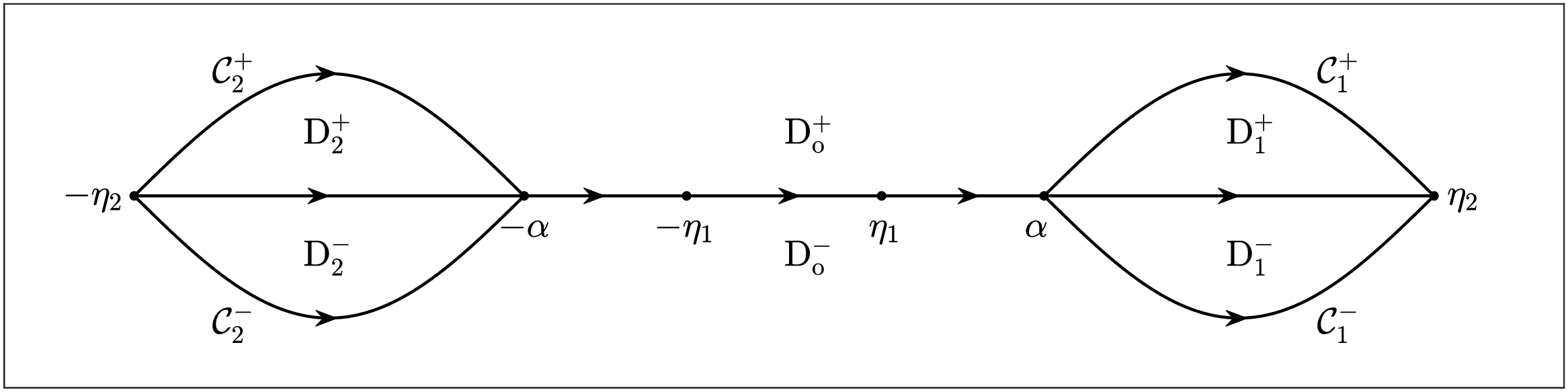}
\caption{Contour deformation by opening lenses in the region $\xi_0<\xi<\eta_2^2$.}
\label{Deformation2}
\end{figure}

Now one can deform the jump contour of the Riemann-Hilbert problem for $T$ by opening lenses, as depicted in Figure \ref{Deformation2}.
The domains $\mathrm{D}^+_1$ and $\mathrm{D}^-_1$ denote lenses  above and below $(\alpha, \eta_2)$, respectively, and
the domains $\mathrm{D}^+_2$ and $\mathrm{D}^-_2$ stand for lenses above and below $(-\eta_2, -\alpha)$, respectively.
Let $\mathrm{D}_j=\mathrm{D}^+_j\cup\mathrm{D}^-_j\, (j=1,2)$, and $\mathrm{D}_\mathrm{o}$ be as the domain outside these lenses, i.e., $\mathrm{D}_\mathrm{o}=\mathbb{C}\setminus \overline{\mathrm{D}_1\cup \mathrm{D}_2\cup(-\alpha, \alpha)}=\mathrm{D}_\mathrm{o}^+\cup \mathrm{D}_\mathrm{o}^- \cup (\eta_2, +\infty)\cup(-\infty, -\eta_2)$,
where $\mathrm{D}_\mathrm{o}^+$ and $\mathrm{D}_\mathrm{o}^-$ denote the parts in the upper and lower half-planes, respectively.
Define the $1\times 2$ vector-valued function $S$ as:
$S=T\mathcal{U}^{tp}_{f}\left[\mathrm{i}r^{-1}\right]^{\mp 1}$, for $\lambda\in\mathrm{D}^{\pm}_1$;
$S=T\mathcal{L}^{tp}_{f}\left[-\mathrm{i}r^{-1}\right]^{\mp 1}$, for $\lambda\in\mathrm{D}^{\pm}_2$;
$S=T$ for $\lambda\in\mathrm{D}_\mathrm{o}$,
where $p$ is defined as
\begin{gather}\label{definition-p}
p=\theta-g.
\end{gather}
As depicted in Fig.~\ref{Deformation2}, $\mathcal{C}^\pm_1$ and $\mathcal{C}^\pm_2$ denote the boundary curves. Define $\mathcal{C}_j=\mathcal{C}^+_j\cup \mathcal{C}^-_j\, (j=1,2)$.
$S$ is analytic in $\lambda$ for $\lambda\in\mathbb{C}\setminus\left(\left[-\eta_2, \eta_2\right]\cup\mathcal{C}_1\cup\mathcal{C}_2\right)$
and
normalizes to
$\begin{pmatrix}
1& 1
\end{pmatrix}$
as $\lambda\to\infty$.
For  $\lambda\in(-\eta_2, \eta_2)\cup\mathcal{C}_1\cup\mathcal{C}_2\setminus\left\{\pm\eta_1, \pm\eta_0, \pm\alpha\right\}$, $S$ admits continuous boundary values, which are related by the following jump conditions
\begin{gather}
S_+=S_-
\begin{cases}
\mathcal{U}^{tp}_{f}\left[\mathrm{i}r^{-1}\right], &\mathrm{for}\,\,\, \lambda\in\mathcal{C}_1,\\[0.5em]
\mathcal{L}^{tp}_{f}\left[-\mathrm{i}r^{-1}\right], &\mathrm{for}\,\,\, \lambda\in\mathcal{C}_2,\\[0.5em]
-\mathrm{i}\sigma_1, &\mathrm{for}\,\,\, \lambda\in\left(\alpha, \eta_2\right),\\[0.5em]
\mathrm{i}\sigma_1, &\mathrm{for}\,\,\, \lambda\in\left(-\eta_2, -\alpha\right),\\[0.5em]
f_-^{-\sigma_3}\mathrm{e}^{tp_-\sigma_3}\mathcal{L}\left[-\mathrm{i}r\right]\mathrm{e}^{-tp_+\sigma_3}f_+^{\sigma_3}, &\mathrm{for}\,\,\, \lambda\in\left(\eta_1, \eta_0\right)\cup\left(\eta_0, \alpha\right),\\[0.5em]
f_-^{-\sigma_3}\mathrm{e}^{tp_-\sigma_3}\mathcal{U}\left[\mathrm{i}r\right]\mathrm{e}^{-tp_+\sigma_3}f_+^{\sigma_3}, &\mathrm{for}\,\,\, \lambda\in\left(-\alpha, -\eta_0\right)\cup\left(-\eta_0, -\eta_1\right),\\[0.5em]
\mathrm{e}^{\left(\Omega t+\Delta\right)\sigma_3}, &\mathrm{for}\,\,\, \lambda\in\left(-\eta_1, \eta_1\right).
\end{cases}
\end{gather}
It is easy to see that $S$ exhibits the same local behaviors near $\pm\eta_1$ and $\pm\eta_0$  as those of $T$, due to $S=T$ for $\lambda\in\mathrm{D}_\mathrm{o}$. This corresponds to the form shown in Eqs.~\eqref{local-Y-eta-1}-\eqref{local-Y-eta-02} with $Y$ replaced by $S$. Near  $\pm\alpha$, the local behavior of $S$ is
$
S=
\mathcal{O}
\begin{pmatrix}
1 &1
\end{pmatrix},
\mathrm{as} \,\,\,\lambda\to\pm\alpha.
$
For the endpoints $\pm\eta_2$, the behaviors of $S$ differ between the regions $\mathrm{D}_{\mathrm{o}}$ and $\mathrm{D}_1\cup \mathrm{D}_2$.
If $\beta_2>0$, $S$ exhibits the following local behavior
\begin{gather}
S=
\begin{cases}
\mathcal{O}
\begin{pmatrix}
 \left|\lambda\mp\eta_2\right|^{\mp\beta_2/2}&\left|\lambda\mp\eta_2\right|^{\pm\beta_2/2}
\end{pmatrix}
, & \mathrm{as} \,\,\,\lambda\in\mathrm{D}_{\mathrm{o}}\to\pm\eta_2, \\[0.5em]
\mathcal{O}
\begin{pmatrix}
 \left|\lambda\mp\eta_2\right|^{-\beta_2/2}&\left|\lambda\mp\eta_2\right|^{-\beta_2/2}
\end{pmatrix}
, & \mathrm{as} \,\,\,\lambda\in\mathrm{D}_1\cup\mathrm{D}_2\to\pm\eta_2,
\end{cases}
\end{gather}
and if $\beta_2\in\left(-1, 0\right)$, the local behavior is formulated as
\begin{gather}
S=
\mathcal{O}
\begin{pmatrix}
 \left|\lambda\mp\eta_2\right|^{\beta_2/2}&\left|\lambda\mp\eta_2\right|^{\beta_2/2}
\end{pmatrix}, \quad
\mathrm{as} \,\,\,\lambda\in\mathrm{D}_1\cup\mathrm{D}_2\cup \mathrm{D}_{\mathrm{o}}\to\pm\eta_2.
\end{gather}

\subsection{Local parametrix near the endpoint $\eta_2$}

For $\lambda\in\left(\mathrm{D}_1\cup\mathrm{D}_\mathrm{o}\right)\cap B(\eta_2)$, the local parametrix $P^{\eta_2}$ can be expressed as
\begin{gather}\label{parametrix-eta_2}
P^{\eta_2}=P^\infty\left(\frac{\mathrm{e}^{\pi\mathrm{i}/4}}{fd}\right)^{\sigma_3}\sigma_1C\zeta_{\eta_2}^{-\sigma_3/4}M^{\mathrm{mB}}\left(\zeta_{\eta_2}, \beta_2\right)\mathrm{e}^{-\sqrt{\zeta_{\eta_2}}\sigma_3}\sigma_1\left(\frac{\mathrm{e}^{\pi\mathrm{i}/4}}{fd}\right)^{-\sigma_3},
\end{gather}
where $\zeta_{\eta_2}=\left(tp\right)^2$ and $P^\infty$ is the outer parametrix in \eqref{outer parametrix}.
The function $d$ is defined as follows:
\begin{itemize}

\item{} For the first generalized reflection coefficient, $r=r_0$, we choose
\bee
d=(\lambda-\eta_1)^{\beta_1/2}(\lambda-\eta_2)^{\beta_2/2}\left|\lambda-\eta_0\right|^{\beta_0/2}\gamma(\lambda)^{1/2};
\ene

\item{} For the second generalized reflection coefficient, $r=r_c$, we take  \bee
    d=(\lambda-\eta_1)^{\beta_1/2}(\lambda-\eta_2)^{\beta_2/2}\chi_c(\lambda)^{1/2}\gamma(\lambda)^{1/2}.
    \ene
\end{itemize}

The matrix $M^{\mathrm{mB}}\left(\zeta_{\eta_2}, \beta_2\right)$ is formulated  using the  modified Bessel functions of the first and second kinds with an index $\beta_2$, namely, $I_{\beta_2}(\zeta_{\eta_2})$ and $K_{\beta_2}(\zeta_{\eta_2})$.
They are two standard solutions of the modified Bessel's equation:
\bee
\zeta_{\eta_2}^2 y''(\zeta_{\eta_2})+\zeta_{\eta_2} y'(\zeta_{\eta_2})-(\zeta_{\eta_2}^2+\beta_2^2)y(\zeta_{\eta_2})=0,
\ene
where $I_{\beta_2}(\zeta_{\eta_2})$ is expressed as
\bee I_{\beta_2}(\zeta_{\eta_2})=(\zeta_{\eta_2}/2)^{\beta_2}\sum_{n=0}^\infty  z^{2n}\left/\right.4^n\Gamma(\beta_2+n+1) n!,
\ene
and the defining property of $K_{\beta_2}(\zeta_{\eta_2})$ is that
$K_{\beta_2}(\zeta_{\eta_2})\sim \sqrt{\pi/2\zeta_{\eta_2}}\mathrm{e}^{-\zeta_{\eta_2}}$, as $\zeta_{\eta_2}\to\infty$ for $\mathrm{arg} \zeta_{\eta_2}\in(-3\pi/2, 3\pi/2) $. Note that the modified Bessel parametrix was first proposed by Kuijlaars, McLaughlin, Assche, and Vanlessen~\cite{61} to match the local behavior near endpoints in their study of strong asymptotics of orthogonal polynomials with respect to the modified Jacobi weight.

In this paper, with a modification, $M^{\mathrm{mB}}(\zeta_{\eta_2}, \beta_2)$ is formulated as follows:
for $\lambda\in\mathrm{D}_\mathrm{o}\cap B(\eta_2)$, the matrix $M^{\mathrm{mB}}(\zeta_{\eta_2}, \beta_2)$ is defined as
\begin{gather} \label{modified-Bessel-1}
M^{\mathrm{mB}}\left(\zeta_{\eta_2}, \beta_2\right)=-\mathrm{i}\sqrt{\pi}
\begin{pmatrix}
\mathrm{i}\sqrt{\zeta_{\eta_2}} & \\[0.5em]
&1
\end{pmatrix}
\begin{pmatrix}
I'_{\beta_2}\left(\sqrt{\zeta_{\eta_2}}\right)  & \mathrm{i}K'_{\beta_2}\left(\sqrt{\zeta_{\eta_2}}\right) \left/\right.\pi\\[0.5em]
I_{\beta_2}\left(\sqrt{\zeta_{\eta_2}}\right)  & \mathrm{i}K_{\beta_2}\left(\sqrt{\zeta_{\eta_2}}\right)\left/\right.\pi
\end{pmatrix};
\end{gather}
for $\lambda\in\mathrm{D}_1^+\cap B(\eta_2)$, it takes the form
\begin{gather}\label{modified-Bessel-2}
M^{\mathrm{mB}}\left(\zeta_{\eta_2}, \beta_2\right)=\frac{1}{\sqrt{\pi}}
\begin{pmatrix}
\mathrm{i}\sqrt{\zeta_{\eta_2}} & \\[1em]
&1
\end{pmatrix}
\begin{pmatrix}
K'_{\beta_2}\left(\sqrt{\zeta_{\eta_2}}\,\mathrm{e}^{-\pi\mathrm{i}}\right)  & K'_{\beta_2}\left(\sqrt{\zeta_{\eta_2}}\right) \\[1em]
-K_{\beta_2}\left(\sqrt{\zeta_{\eta_2}}\,\mathrm{e}^{-\pi\mathrm{i}}\right)  & K_{\beta_2}\left(\sqrt{\zeta_{\eta_2}}\right)
\end{pmatrix};
\end{gather}
for $\lambda\in\mathrm{D}_1^-\cap B(\eta_2)$, it is expressed as
\begin{gather}\label{modified-Bessel-3}
M^{\mathrm{mB}}\left(\zeta_{\eta_2}, \beta_2\right)=\frac{1}{\sqrt{\pi}}
\begin{pmatrix}
\mathrm{i}\sqrt{\zeta_{\eta_2}} & \\[0.5em]
&1
\end{pmatrix}
\begin{pmatrix}
-K'_{\beta_2}\left(\sqrt{\zeta_{\eta_2}}\,\mathrm{e}^{\pi\mathrm{i}}\right)  & K'_{\beta_2}\left(\sqrt{\zeta_{\eta_2}}\right) \\[1em]
K_{\beta_2}\left(\sqrt{\zeta_{\eta_2}}\,\mathrm{e}^{\pi\mathrm{i}}\right)  & K_{\beta_2}\left(\sqrt{\zeta_{\eta_2}}\right)
\end{pmatrix}.
\end{gather}
The matrix $M^{\mathrm{mB}}\left(\zeta_{\eta_2}, \beta_2\right)$ addresses  a RH problem characterized by specific properties, as depicted in  Fig.~\ref{BesselAiry}(left).
$M^{\mathrm{mB}}\left(\zeta_{\eta_2}, \beta_2\right)$ is analytic in $\zeta_{\eta_2}$ for $\zeta_{\eta_2}\in\mathbb{C}\setminus(\Sigma_1\cup\Sigma_2\cup\Sigma_3)$,
and as $\zeta_{\eta_2}\to\infty$, it normalizes to
\begin{gather}
M^{\mathrm{mB}}\left(\zeta_{\eta_2}, \beta_2\right)=\zeta_{\eta_2}^{\sigma_3/4}C^{-1}\left(\mathbb{I}_2+\mathcal{O}\left(\frac{1}{\sqrt{\zeta_{\eta_2}}}\right)\right)\mathrm{e}^{\sqrt{\zeta_{\eta_2}}\,\sigma_3}.
\end{gather}
For $\zeta_{\eta_2}\in\Sigma_1^0\cup\Sigma_2^0\cup\Sigma_3^0$ with $\Sigma_j^0=\Sigma_j\setminus \{0\}, \, j=1, 2, 3$,  $M^{\mathrm{mB}}\left(\zeta_{\eta_2}, \beta_2\right)$ admits continuous boundary values, and they are related by the following jump conditions
\begin{gather}
M^{\mathrm{mB}}_+\left(\zeta_{\eta_2}, \beta_2\right)=M^{\mathrm{mB}}_-\left(\zeta_{\eta_2}, \beta_2\right)
\begin{cases}
\mathcal{L}\left[\mathrm{e}^{\beta_2 \pi \mathrm{i}}\right], &\mathrm{for}\,\,\, \zeta_{\eta_2}\in\Sigma_1^0, \\[0.5em]
\mathrm{i}\sigma_2, &\mathrm{for}\,\,\, \zeta_{\eta_2}\in\Sigma_2^0, \\[0.5em]
\mathcal{L}\left[\mathrm{e}^{-\beta_2 \pi \mathrm{i}}\right], &\mathrm{for}\,\,\, \zeta_{\eta_2}\in\Sigma_3^0.
\end{cases}
\end{gather}
Near the origin in the $\zeta_{\eta_2}$-plane, $M^{\mathrm{mB}}\left(\zeta_{\eta_2}, \beta_2\right)$ exhibits the following local behavior. If $\beta_2>0$, the local behavior is
\begin{gather}
M^{\mathrm{mB}}\left(\zeta_{\eta_2}, \beta_2\right)=
\begin{cases}
\mathcal{O}
\begin{pmatrix}
\left|\zeta_{\eta_2}\right|^{\beta_2/2} & \left|\zeta_{\eta_2}\right|^{-\beta_2/2}  \\[1em]
\left|\zeta_{\eta_2}\right|^{\beta_2/2} & \left|\zeta_{\eta_2}\right|^{-\beta_2/2}
\end{pmatrix},
& \mathrm{as}\,\,\, \zeta_{\eta_2}\in \mathrm{D}^\zeta_1\to 0, \\[3em]
\mathcal{O}
\begin{pmatrix}
\left|\zeta_{\eta_2}\right|^{-\beta_2/2} & \left|\zeta_{\eta_2}\right|^{-\beta_2/2}  \\[1em]
\left|\zeta_{\eta_2}\right|^{-\beta_2/2} & \left|\zeta_{\eta_2}\right|^{-\beta_2/2}
\end{pmatrix},
& \mathrm{as}\,\,\, \zeta_{\eta_2}\in \mathrm{D}^\zeta_2\cup\mathrm{D}^\zeta_3\to 0,
\end{cases}
\end{gather}
if $\beta_2=0$, the local behavior is exhibited as
\begin{gather}
M^{\mathrm{mB}}\left(\zeta_{\eta_2}, \beta_2=0\right)=
\begin{cases}
\mathcal{O}
\begin{pmatrix}
1 & \log\left|\zeta_{\eta_2}\right|  \\[0.5em]
1 & \log\left|\zeta_{\eta_2}\right|
\end{pmatrix},
& \mathrm{as}\,\,\, \zeta_{\eta_2}\in \mathrm{D}^\zeta_1\to 0, \\[2em]
\mathcal{O}
\begin{pmatrix}
\log\left|\zeta_{\eta_2}\right| & \log\left|\zeta_{\eta_2}\right|  \\[0.5em]
\log\left|\zeta_{\eta_2}\right| & \log\left|\zeta_{\eta_2}\right|
\end{pmatrix},
& \mathrm{as}\,\,\, \zeta_{\eta_2}\in \mathrm{D}^\zeta_2\cup\mathrm{D}^\zeta_3\to 0,
\end{cases}
\end{gather}
and if $-1<\beta_2<0$, $M^{\mathrm{mB}}\left(\zeta_{\eta_2}, \beta_2\right)$ exhibits the local behavior
\begin{gather}
M^{\mathrm{mB}}\left(\zeta_{\eta_2}, \beta_2\right)=\mathcal{O}
\begin{pmatrix}
\left|\zeta_{\eta_2}\right|^{\beta_2/2} & \left|\zeta_{\eta_2}\right|^{\beta_2/2}  \\[0.5em]
\left|\zeta_{\eta_2}\right|^{\beta_2/2} & \left|\zeta_{\eta_2}\right|^{\beta_2/2}
\end{pmatrix}, \quad
\mathrm{as}\,\,\, \zeta_{\eta_2}\in \mathrm{D}^\zeta_1\cup\mathrm{D}^\zeta_2\cup\mathrm{D}^\zeta_3\to 0.
\end{gather}

\begin{figure}[!t]
\centering
\includegraphics[scale=0.25]{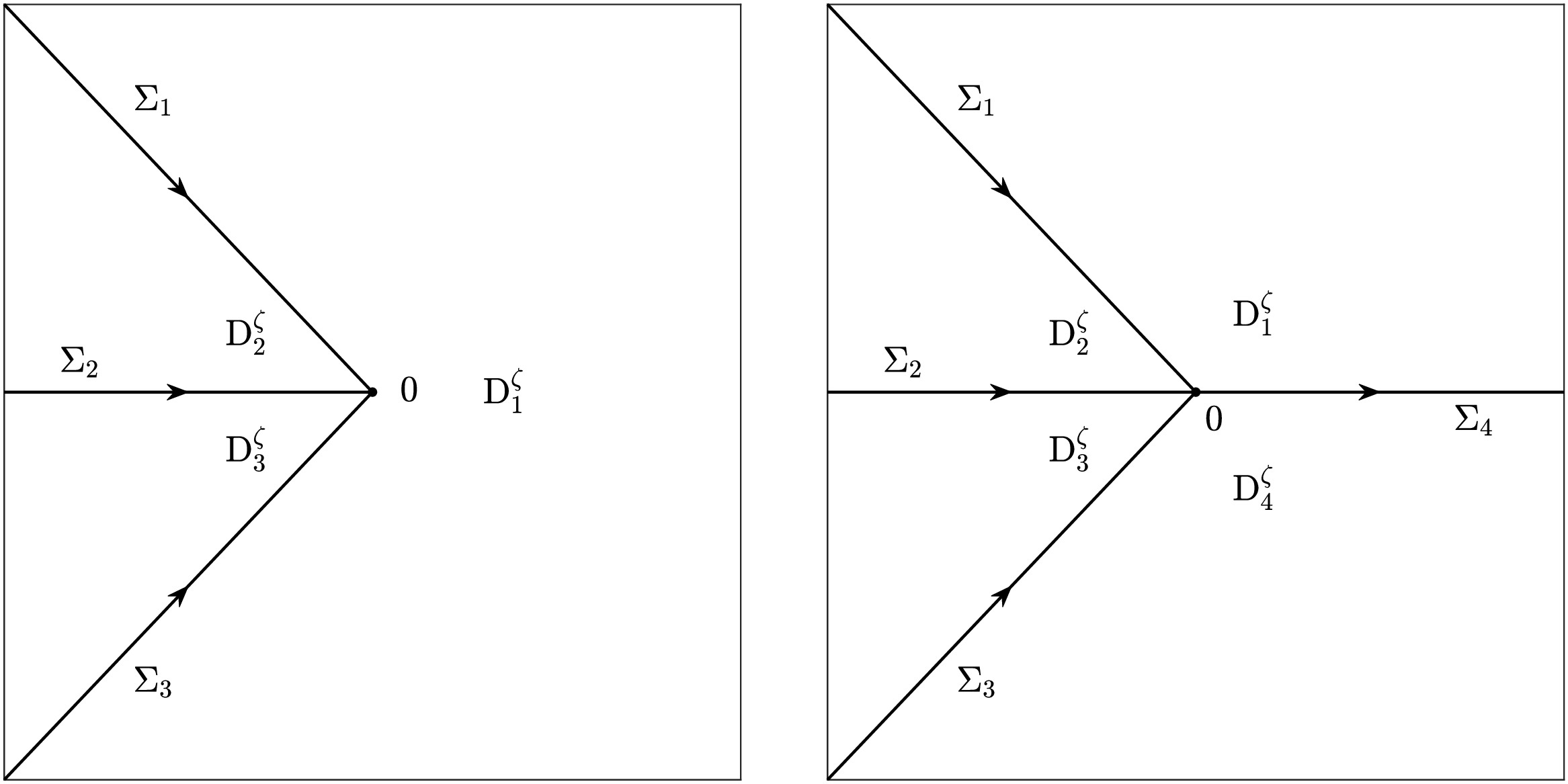}
\caption{Left: Jump contours for modified Bessel parametrix $M^{\mathrm{mB}}$; Right: Jump contours for the Airy parametrix $M^{\mathrm{Ai}}$.}
\label{BesselAiry}
\end{figure}

It follows from the above that the matrix $P^{\eta_2}$ satisfies a RH problem in the neighborhood of $\lambda=\eta_2$ characterized by specific properties.
$P^{\eta_2}$ is analytic in $\lambda$ for $\lambda\in\left(\mathrm{D}_1\cup\mathrm{D}_\mathrm{o}\right)\cap B(\eta_2)$,
and as $t\to +\infty$, it normalizes to
\begin{gather}\label{small-norm-eta_2}
P^{\eta_2}\left(P^\infty\right)^{-1}=\mathbb{I}_2+\mathcal{O}\left(\frac{1}{t}\right),
\end{gather}
which  holds uniformly for $\lambda\in\partial B\left(\eta_2\right)$.
For $\lambda\in\big(\mathcal{C}_1\cup (\alpha, \eta_2)\big)\cap B(\eta_2)$, $P^{\eta_2}$ admits continuous boundary values, denoted as $P^{\eta_2}_+$ and $P^{\eta_2}_-$. These values are related by the following jump relations
\begin{gather}
P^{\eta_2}_+=P^{\eta_2}_-
\begin{cases}
\mathcal{U}^{tp}_{f}\left[\mathrm{i}r^{-1}\right], &\mathrm{for}\,\,\, \lambda\in\mathcal{C}_1\cap B(\eta_2),\\[0.5em]
-\mathrm{i}\sigma_1, &\mathrm{for}\,\,\, \lambda\in\left(\alpha, \eta_2\right)\cap B(\eta_2).
\end{cases}
\end{gather}
If $\beta_2>0$, $P^{\eta_2}$ exhibits the following local behavior
\begin{gather}
P^{\eta_2}=
\begin{cases}
\mathcal{O}
\begin{pmatrix}
 \left|\lambda-\eta_2\right|^{-\beta_2/2}&\left|\lambda-\eta_2\right|^{\beta_2/2}  \\[1em]
  \left|\lambda-\eta_2\right|^{-\beta_2/2}&\left|\lambda-\eta_2\right|^{\beta_2/2}
\end{pmatrix}
, & \mathrm{as} \,\,\,\lambda\in\mathrm{D}_\mathrm{o}\to\eta_2, \\[2em]
\mathcal{O}
\begin{pmatrix}
 \left|\lambda-\eta_2\right|^{-\beta_2/2}&\left|\lambda-\eta_2\right|^{-\beta_2/2}  \\[1em]
  \left|\lambda-\eta_2\right|^{-\beta_2/2}&\left|\lambda-\eta_2\right|^{-\beta_2/2}
\end{pmatrix}
, & \mathrm{as} \,\,\,\lambda\in\mathrm{D}_1\to\eta_2,
\end{cases}
\end{gather}
and if $\beta_2\in\left(-1, 0\right)$, the local behavior is formulated as
\begin{gather}
P^{\eta_2}=
\mathcal{O}
\begin{pmatrix}
 \left|\lambda-\eta_2\right|^{\beta_2/2}&\left|\lambda-\eta_2\right|^{\beta_2/2}\\[1em]
  \left|\lambda-\eta_2\right|^{\beta_2/2}&\left|\lambda-\eta_2\right|^{\beta_2/2}
\end{pmatrix}, \quad
\mathrm{as} \,\,\,\lambda\in\mathrm{D}_1\cup\mathrm{D}_\mathrm{o}\to\eta_2.
\end{gather}

\subsection{Local parametrix near $\lambda=\alpha$}

The local parametrix in the neighbourhood of $\lambda=\alpha$, denoted as $P^{\alpha}$, is expressed as follows:
\begin{itemize}

\item{} For $\lambda\in\left(\mathrm{D}_1^+\cup\mathrm{D}_\mathrm{o}^+\right)\cap B(\alpha)$, $P^{\alpha}$ is written as
\begin{gather}\label{parametrix-alpha-1}
P^{\alpha}=P^\infty\left(\frac{\mathrm{e}^{\pi\mathrm{i}/4}-\Omega t/2}{f\sqrt{r}}\right)^{\sigma_3}\sigma_2C\zeta_{\alpha}^{-\sigma_3/4}M^{\mathrm{Ai}}(\zeta_{\alpha})\mathrm{e}^{\zeta_{\alpha}^{3/2}\sigma_3}\sigma_2\left(\frac{\mathrm{e}^{\pi\mathrm{i}/4}-\Omega t/2}{f\sqrt{r}}\right)^{-\sigma_3},
\end{gather}
where $\zeta_{\alpha}=t^{2/3}\left(p+\Omega/2\right)^{2/3}$;

\item{} For $\lambda\in\left(\mathrm{D}_\mathrm{o}^-\cup\mathrm{D}_1^-\right)\cap B(\alpha)$, $P^{\alpha}$ is written as
\begin{gather}\label{parametrix-alpha-2}
P^{\alpha}=P^\infty\left(\frac{\mathrm{e}^{\pi\mathrm{i}/4}+\Omega t/2}{f\sqrt{r}}\right)^{\sigma_3}\sigma_2C\zeta_{\alpha}^{-\sigma_3/4}M^{\mathrm{Ai}}(\zeta_{\alpha})\mathrm{e}^{\zeta_{\alpha}^{3/2}\sigma_3}\sigma_2\left(\frac{\mathrm{e}^{\pi\mathrm{i}/4}+\Omega t/2}{f\sqrt{r}}\right)^{-\sigma_3},
\end{gather}
where
$\zeta_{\alpha}=t^{2/3}\left(p-\Omega/2\right)^{2/3}$.
\end{itemize}
The matrix $M^{\mathrm{Ai}}(\zeta_\alpha)$ is constructed using the Airy function of the first kind, $\mathrm{Ai}(\zeta_\alpha)$, which is the standard solution of the Airy's equation $y''(\zeta_\alpha)-\zeta_\alpha y(\zeta_\alpha)=0$. Note that
the Airy parametrix was first introduced by Deift, Kriecherbauer, McLaughlin, Venakides, and Zhou~\cite{49} in the study of the strong asymptotics of orthogonal polynomials with respect to exponential weights using the Deift-Zhou steepest descent method  for a RH problem \cite{51}.

In this paper, with a modification, the matrix $M^{\mathrm{Ai}}(\zeta_\alpha)$ is formulated as follows:
for $\lambda\in\mathrm{D}_\mathrm{o}^-\cap B(\alpha)$, the matrix $M^{\mathrm{Ai}}\left(\zeta_\alpha\right)$ is in the form
\begin{gather}
M^{\mathrm{Ai}}\left(\zeta_\alpha\right)=\zeta_0^{-\sigma_3/4}C_0
\begin{pmatrix}
\mathrm{Ai}'\left(\zeta_0\zeta_\alpha\right)  & -\mathrm{e}^{2\pi \mathrm{i}/3} \mathrm{Ai}'\left(\mathrm{e}^{-2\pi\mathrm{i}/3}\zeta_0\zeta_\alpha\right) \\[0.5em]
\mathrm{Ai}\left(\zeta_0\zeta_\alpha\right)  & -\mathrm{e}^{-2\pi\mathrm{i}/3} \mathrm{Ai}\left(\mathrm{e}^{-2\pi\mathrm{i}/3}\zeta_0\zeta_\alpha\right)
\end{pmatrix};
\end{gather}
for $\zeta\in\mathrm{D}_1^-\cap B(\alpha)$, the matrix $M^{\mathrm{Ai}}\left(\zeta_\alpha\right)$ is formulated as
\begin{gather}
M^{\mathrm{Ai}}\left(\zeta_\alpha\right)=\zeta_0^{-\sigma_3/4}C_0
\begin{pmatrix}
-\mathrm{e}^{2\pi \mathrm{i}/3}\mathrm{Ai}'\left(\mathrm{e}^{-4\pi \mathrm{i}/3}\zeta_0\zeta_\alpha\right)  & -\mathrm{e}^{-4\pi \mathrm{i}/3} \mathrm{Ai}'\left(\mathrm{e}^{2\pi \mathrm{i}/3}\zeta_0\zeta_\alpha\right) \\[0.5em]
-\mathrm{e}^{-4\pi \mathrm{i}/3}\mathrm{Ai}\left(\mathrm{e}^{-4\pi \mathrm{i}/3}\zeta_0\zeta_\alpha\right)  & -\mathrm{e}^{2\pi \mathrm{i}/3} \mathrm{Ai}\left(\mathrm{e}^{2\pi \mathrm{i}/3}\zeta_0\zeta_\alpha\right)
\end{pmatrix};
\end{gather}
for $\zeta\in\mathrm{D}_1^+\cap B(\alpha)$, the matrix $M^{\mathrm{Ai}}\left(\zeta_\alpha\right)$ is written as
\begin{gather}
M^{\mathrm{Ai}}\left(\zeta_\alpha\right)=\zeta_0^{-\sigma_3/4}C_0
\begin{pmatrix}
-\mathrm{e}^{2\pi \mathrm{i}/3}\mathrm{Ai}'\left(\mathrm{e}^{4\pi \mathrm{i}/3}\zeta_0\zeta_\alpha\right)  & \mathrm{e}^{4\pi \mathrm{i}/3} \mathrm{Ai}'\left(\mathrm{e}^{2\pi \mathrm{i}/3}\zeta_0\zeta_\alpha\right) \\[0.5em]
-\mathrm{e}^{4\pi \mathrm{i}/3}\mathrm{Ai}\left(\mathrm{e}^{4\pi \mathrm{i}/3}\zeta_0\zeta_\alpha\right)  & \mathrm{e}^{2\pi \mathrm{i}/3} \mathrm{Ai}\left(\mathrm{e}^{2\pi \mathrm{i}/3}\zeta_0\zeta_\alpha\right)
\end{pmatrix};
\end{gather}
for $\zeta\in\mathrm{D}_\mathrm{o}^+\cap B(\alpha)$, the matrix $M^{\mathrm{Ai}}\left(\zeta_\alpha\right)$ is expressed as
\begin{gather}
M^{\mathrm{Ai}}\left(\zeta_\alpha\right)=\zeta_0^{-\sigma_3/4}C_0
\begin{pmatrix}
\mathrm{Ai}'\left(\zeta_0\zeta_\alpha\right)  & \mathrm{e}^{4\pi \mathrm{i}/3} \mathrm{Ai}'\left(\mathrm{e}^{2\pi \mathrm{i}/3}\zeta_0\zeta_\alpha\right) \\[0.5em]
\mathrm{Ai}\left(\zeta_0\zeta_\alpha\right)  & \mathrm{e}^{2\pi \mathrm{i}/3} \mathrm{Ai}\left(\mathrm{e}^{2\pi \mathrm{i}/3}\zeta_0\zeta_\alpha\right)
\end{pmatrix}.
\end{gather}
In the above formulas $\zeta_0=\left(2/3\right)^{-2/3}$.
$M^{\mathrm{Ai}}\left(\zeta_\alpha\right)$ solves a Riemann-Hilbert problem with the following properties, as depicted in Figure \ref{BesselAiry}(Right). $M^{\mathrm{Ai}}\left(\zeta_\alpha\right)$ is analytic in $\zeta_\alpha$ for $\zeta_\alpha\in\mathbb{C}\setminus (\cup_{j=1}^4\Sigma_j)$, and as $\zeta_\alpha\to\infty$, it normalizes to
\begin{gather}
M^{\mathrm{Ai}}\left(\zeta_\alpha\right)=\zeta_\alpha^{\sigma_3/4}C^{-1}
\left(\mathbb{I}_2+\mathcal{O}\left(\zeta_\alpha^{-3/2}\right)\right)\mathrm{e}^{-\zeta_\alpha^{3/2}\,\sigma_3}.
\end{gather}
For $\zeta_\alpha\in\cup_{j=1}^4\Sigma_j^0$ with $\Sigma_j^0=\Sigma_j\setminus \{0\}, \, j=1, 2, 3, 4$,  $M^{\mathrm{Ai}}\left(\zeta_\alpha\right)$ admits continuous boundary values, and they are related by the following jump conditions
\begin{gather}
M^{\mathrm{Ai}}_+\left(\zeta_\alpha\right)=M^{\mathrm{Ai}}_-\left(\zeta_\alpha\right)
\begin{cases}
\mathcal{L}\left[1\right], &\mathrm{for}\,\,\, \zeta_\alpha\in\Sigma_1^0\cup\Sigma_3^0, \\[0.5em]
\mathrm{i}\sigma_2, &\mathrm{for}\,\,\, \zeta_\alpha\in\Sigma_2^0, \\[0.5em]
\mathcal{U}\left[1\right], &\mathrm{for}\,\,\, \zeta_\alpha\in\Sigma_3^0,
\end{cases}
\end{gather}
 Near the self-intersection point $\zeta_\alpha=0$,  the $M^{\mathrm{Ai}}\left(\zeta_\alpha\right)$ exhibits the following local behavior
\begin{gather}
M^{\mathrm{Ai}}\left(\zeta_\alpha\right)=
\mathcal{O}
\begin{pmatrix}
1 & 1\\
1 & 1
\end{pmatrix},
\quad \mathrm{as}\,\,\, \zeta_\alpha\to 0.
\end{gather}
It follows from the above that the matrix $P^{\alpha}$ satisfies a Riemann-Hilbert problem in the neighborhood of $\lambda=\alpha$, characterized by specific properties.
$P^{\alpha}$ is analytic in $\lambda$ for $\lambda\in\left(\mathrm{D}_1\cup\mathrm{D}_\mathrm{o}\right)\cap B(\alpha)$,
and as $t\to +\infty$, it normalizes to
\begin{gather}\label{small-norm-alpha}
P^{\alpha}\left(P^\infty\right)^{-1}=\mathbb{I}_2+\mathcal{O}\left(\frac{1}{t}\right),
\end{gather}
which  holds uniformly for $\lambda\in\partial B\left(\alpha\right)$.
For $\lambda\in\big(\mathcal{C}_1\cup (\alpha, \eta_2)\big)\cap B(\eta_2)$, $P^{\eta_2}$ admits continuous boundary values, denoted as $P^{\eta_2}_+$ and $P^{\eta_2}_-$. These values are related by the following jump relations
\begin{gather}
P^{\alpha}_+=P^{\alpha}_-
\begin{cases}
\mathcal{U}^{tp}_{f}\left[\mathrm{i}r^{-1}\right], &\mathrm{for}\,\,\, \lambda\in\mathcal{C}_1\cap B(\alpha),\\[0.5em]
-\mathrm{i}\sigma_1, &\mathrm{for}\,\,\, \lambda\in\left(\alpha, \eta_2\right)\cap B(\alpha),\\[0.5em]
f_-^{-\sigma_3}\mathrm{e}^{tp_-\sigma_3}\mathcal{L}\left[-\mathrm{i}r\right]\mathrm{e}^{-tp_+\sigma_3}f_+^{\sigma_3}, & \mathrm{for}\,\,\, \lambda\in\left(\eta_0, \alpha\right)\cap B(\alpha).
\end{cases}
\end{gather}
Near  $\alpha$, the matrix $P^{\alpha}$ exhibits the following behavior
\begin{gather}
P^{\alpha}=
\mathcal{O}
\begin{pmatrix}
1 & 1\\
1 & 1
\end{pmatrix},
\quad \mathrm{as}\,\,\, \lambda\in\mathrm{D}_1\cup\mathrm{D}_\mathrm{o}\to\alpha.
\end{gather}

\subsection{Riemann-Hilbert problem for $E$}

The error vector $E$ is defined by
\begin{gather}\label{error vector}
E=SP^{-1},
\end{gather}
where $P$ is the global parametrix, formulated as
\begin{gather}\label{global parametrix}
P=
\begin{cases}
P^\infty\left(\lambda\right), &\mathrm{for}\,\,\,\lambda\in\mathbb{C}\setminus\overline{B\left(\pm\eta_2, \pm\alpha\right)},  \v\\
P^{\eta_2}\left(\lambda\right), & \mathrm{for}\,\,\,\lambda\in B\left(\eta_2\right),  \v\\
P^{\alpha}\left(\lambda\right), & \mathrm{for}\,\,\,\lambda\in B\left(\alpha\right),  \v\\
\sigma_1P^{\eta_2}\left(-\lambda\right)\sigma_1, & \mathrm{for}\,\,\,\lambda\in B\left(-\eta_2\right),  \v\\
\sigma_1P^{\alpha}\left(-\lambda\right)\sigma_1, & \mathrm{for}\,\,\,\lambda\in B\left(-\alpha\right)
\end{cases}
\end{gather}
with $B\left(\pm\eta_2, \pm\alpha\right)=B(\eta_2)\cup B(-\eta_2)\cup B(\alpha)\cup B(-\alpha)$.
Note that $P^\infty$ is the same as that in  \cite{15}, which is included here for completeness. $P^\infty$ takes the form of a  $2\times 2$ matrix
\begin{gather}\label{outer parametrix}
P^\infty=
\begin{pmatrix}
1 &1 \\[1em]
\lambda&-\lambda
\end{pmatrix}^{-1}
\begin{pmatrix}
S^\infty  \\[1em]
S^\infty_x+\left(tp\right)_xS^\infty\sigma_3
\end{pmatrix},
\end{gather}
where $S^\infty$ is a $1\times 2$ vector, written as  $S^\infty=\left(S^\infty_1 \,\,\, S^\infty_2\right)$,
\begin{gather}
\begin{aligned}
S^\infty_1&=\left(\frac{\lambda^2-\alpha^2}{\lambda^2-\eta_2^2}\right)^{1/4}\frac{\vartheta_3\left(2w-\frac{1}{2}+\frac{\Omega t+\Delta}{2\pi\mathrm{i}}\right)}{\vartheta_3\left(2w-\frac{1}{2}; 2\tau\right)}\frac{\vartheta_3\left(0; 2\tau\right)}{\vartheta_3\left(\frac{\Omega t+\Delta}{2\pi\mathrm{i}}; 2\tau\right)}, \\[0.5em]
S^\infty_2&=\left(\frac{\lambda^2-\alpha^2}{\lambda^2-\eta_2^2}\right)^{1/4}\frac{\vartheta_3\left(-2w-\frac{1}{2}+\frac{\Omega t+\Delta}{2\pi\mathrm{i}}\right)}{\vartheta_3\left(-2w-\frac{1}{2}; 2\tau\right)}\frac{\vartheta_3\left(0; 2\tau\right)}{\vartheta_3\left(\frac{\Omega t+\Delta}{2\pi\mathrm{i}}; 2\tau\right)},
\end{aligned}
\end{gather}
with
\bee
\vartheta_3(\lambda; \tau)=\sum_{n\in\mathbb{Z}}\exp(2\pi\mathrm{i}n\lambda+\pi\mathrm{i}n^2\tau),
\quad
\tau=\mathrm{i}eK(\sqrt{1-m^2})\left/\right.(2eK(m))\in\mathrm{i}\mathbb{R},
\ene and
\bee
w=\d\dfrac{\d\int_{\eta_2}^\lambda R^{-1}(s)\mathrm{d}s}{\d 4\int_0^\alpha R^{-1}(s)\mathrm{d}s}.
\ene
By utilizing the symmetry $P^\infty\left(\lambda\right)=\sigma_1P^\infty\left(-\lambda\right)\sigma_1$ and the standard small norm argument, the soliton gas exhibits the asymptotic behavior given by \eqref{asymptotic-middle} in the region $\xi\in\left(\xi_0, \eta_2^2\right)$. It is evident from the asymptotic behavior that the error decays at a rate of $\mathcal{O}\left(t^{-1}\right)$.
This low decay rate is also observed in the study of the infinite-order rogue wave for the NLS equation \cite{33}.

 \section{Long-time asymptotics: the region $\xi_{\mathrm{crit}}<\xi<\xi_0$}

The Riemann-Hilbert problem for $Y$ was described in the previous section. It is important to note that in this section, $\eta_1 < \alpha < \eta_0$, whereas  in the previous section, $\eta_0 < \alpha < \eta_2$ for the region $\xi_0 < \xi < \eta_2^2$.
The primary significant difference in the Deift-Zhou steepest descent technique between the regions $\xi_0 < \xi < \eta_2^2$ and $\xi_{\mathrm{crit}} < \xi < \xi_0$ lies in the Riemann-Hilbert problem for $T$.

\subsection{Riemann-Hilbert problems for $T$ and $S$}
$T$ is still defined by the conjugation \eqref{Conjugation} with the $g$-function \eqref{g-function} and the $f$-function \eqref{f-function}.
$T$ is analytic in $\lambda$ for $\lambda\in\mathbb{C}\setminus\left(\left[-\eta_2, \eta_2\right]\right)$ and normalizes to $
\begin{pmatrix}
1& 1
\end{pmatrix}$
as $\lambda\to\infty$.
Compared to the region $\xi_0<\xi<\eta_2^2$, the jump conditions exhibit some differences.
For  $\lambda\in\left(-\eta_2, \eta_2\right)\setminus\left\{\pm\eta_1, \pm\eta_0, \pm\alpha\right\}$, $T$ admits continuous boundary values, which are related by the following jump conditions
\begin{gather}
T_+=T_-
\begin{cases}
\mathcal{U}^{tp_-}_{f_-}\left[\mathrm{i}r^{-1}\right]\left(-\mathrm{i}\sigma_1\right)\mathcal{U}^{tp_+}_{f_+}\left[\mathrm{i}r^{-1}\right], &\mathrm{for}\,\,\, \lambda\in\left(\alpha, \eta_0\right)\cup\left(\eta_0, \eta_2\right),\\[0.5em]
\mathcal{L}^{tp_-}_{f_-}\left[-\mathrm{i}r^{-1}\right]\left(\mathrm{i}\sigma_1\right)\mathcal{L}^{tp_+}_{f_+}\left[-\mathrm{i}r^{-1}\right], &\mathrm{for}\,\,\, \lambda\in\left(-\eta_2, -\eta_0\right)\cup\left(-\eta_0, -\alpha\right),\\[0.5em]
f_-^{-\sigma_3}\mathrm{e}^{tp_-\sigma_3}\mathcal{L}\left[-\mathrm{i}r\right]\mathrm{e}^{-tp_+\sigma_3}f_+^{\sigma_3}, &\mathrm{for}\,\,\, \lambda\in\left(\eta_1, \alpha\right),\\[0.5em]
f_-^{-\sigma_3}\mathrm{e}^{tp_-\sigma_3}\mathcal{U}\left[\mathrm{i}r\right]\mathrm{e}^{-tp_+\sigma_3}f_+^{\sigma_3}, &\mathrm{for}\,\,\, \lambda\in\left(-\alpha, -\eta_1\right),\\[0.5em]
\mathrm{e}^{\left(\Omega t+\Delta\right)\sigma_3}, &\mathrm{for}\,\,\, \lambda\in\left(-\eta_1, \eta_1\right).
\end{cases}
\end{gather}
The local behaviors of $T$ near $\pm\eta_1$, $\pm\eta_2$, and $\pm\alpha$ are the same as those in the region $\xi_0<\xi<\eta_2^2$.
$T$ also has the same local behavior near $\pm\eta_0$ for the second generalized reflection coefficient $r_c$.
However, for the first  generalized reflection coefficient $r_0$, $T$ exhibits the following local behaviors
\begin{gather}
T=
\begin{cases}
\mathcal{O}
\begin{pmatrix}
 \left|\lambda\mp\eta_0\right|^{\mp\beta_0/2}&\left|\lambda\mp\eta_0\right|^{\pm\beta_0/2}
\end{pmatrix}
, & \mathrm{if} \,\,\,\beta_0\in\left(0, +\infty\right), \\[1em]
\mathcal{O}
\begin{pmatrix}
 \left|\lambda\mp\eta_0\right|^{\beta_0/2}&\left|\lambda\mp\eta_0\right|^{\beta_0/2}
\end{pmatrix}
, & \mathrm{if} \,\,\,\beta_0\in\left(-1, 0\right).
\end{cases}
\end{gather}

\begin{figure}[!t]
\centering
\includegraphics[scale=0.35]{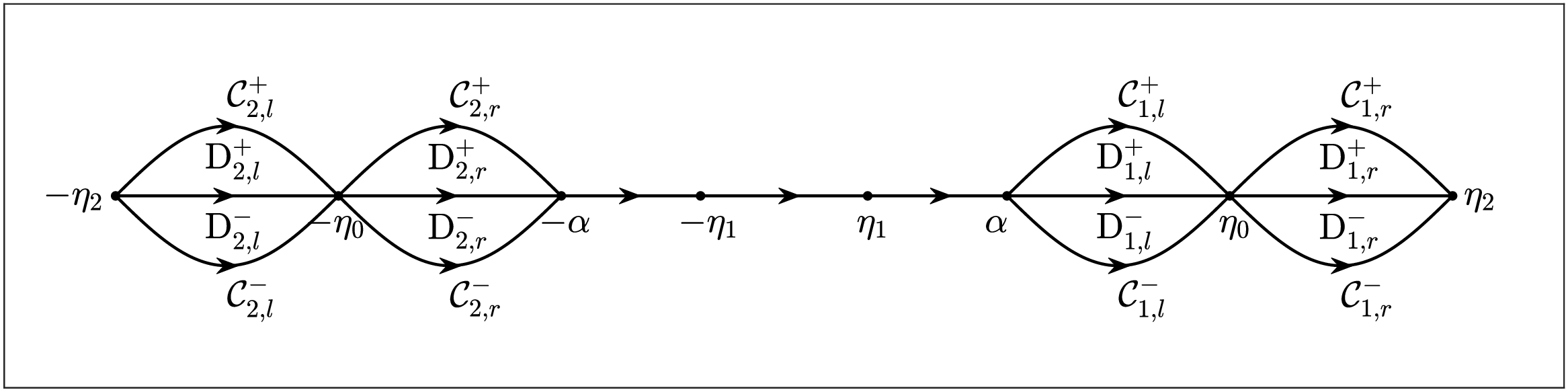}
\caption{Contour deformation by opening lenses in the region $\xi_{\mathrm{crit}}<\xi<\xi_0$.}
\label{Deformation3}
\end{figure}

Since $r_0$ and $r_c$ are not analytic at $\lambda=\pm\eta_0$, one must open lenses above and below $\left(\alpha, \eta_0\right)$, $\left(\eta_0, \eta_2\right)$, $\left(-\eta_2, -\eta_0\right)$, and $\left(-\eta_0, -\alpha\right)$ in the region $\xi_{\mathrm{crit}}<\xi<\xi_0$ rather than $\left(\alpha, \eta_2\right)$ and $\left(-\eta_2, -\alpha\right)$ in the region $\xi_0<\xi<\eta_2^2$.
The contour deformation of the Riemann-Hilbert problem by opening lenses is depicted in Figure \ref{Deformation3}.
The domains $\mathrm{D}^+_{1, l}$ and $\mathrm{D}^-_{1, l}$ are lenses, respectively, above and below $(\alpha, \eta_0)$.
The domains $\mathrm{D}^+_{1, r}$ and $\mathrm{D}^-_{1, r}$ are lenses, respectively, above and below $(\eta_0, \eta_2)$.
The domains $\mathrm{D}^+_{2, l}$ and $\mathrm{D}^-_{2, l}$ are lenses, respectively, above and below $(-\eta_2, -\eta_0)$.
The domains $\mathrm{D}^+_{2, r}$ and $\mathrm{D}^-_{2, r}$ are lenses, respectively, above and below $(-\eta_0, -\alpha)$.
For convenience, define
$\mathrm{D}_{j, l}=\mathrm{D}_{j, l}^+\cup \mathrm{D}_{j, l}^-$,
$\mathrm{D}_{j, r}=\mathrm{D}_{j, r}^+\cup \mathrm{D}_{j, r}^-$,
$\mathrm{D}_j^{\pm}=\mathrm{D}_{j, l}^{\pm}\cup \mathrm{D}_{j, r}^{\pm}$,
and $\mathrm{D}_j=\mathrm{D}_{j, l}\cup \mathrm{D}_{j, r}$,\, $j=1,2$.
Denote $\mathrm{D}_\mathrm{o}$ as the domain outside these lenses, i.e.,
$\mathrm{D}_\mathrm{o}=\mathbb{C}\setminus \overline{\mathrm{D}_1\cup \mathrm{D}_2\cup(-\alpha, \alpha)}$
and $\mathrm{D}_\mathrm{o}=\mathrm{D}_\mathrm{o}^+\cup \mathrm{D}_\mathrm{o}^- \cup (\eta_2, +\infty)\cup(-\infty, -\eta_2)$,
where $\mathrm{D}_\mathrm{o}^+$ and $\mathrm{D}_\mathrm{o}^-$ denote the parts in the upper half-plane and lower half-plane, respectively.
The $1\times 2$ vector-valued function $S$ are defined as follows:
$S=T\mathcal{U}^{tp}_{f}\left[\mathrm{i}r^{-1}\right]^{\mp 1}$ for $\lambda\in\mathrm{D}_1^{\pm}$;
$S=T\mathcal{L}^{tp}_{f}\left[-\mathrm{i}r^{-1}\right]^{\mp 1}$ for $\lambda\in\mathrm{D}_2^{\pm}$;
$S=T$ for $\lambda\in\mathrm{D}_\mathrm{o}$,
where $p$ is still defined in \eqref{definition-p}.
The defining function $S$ satisfies the following Riemann-Hilbert problem.
$S$ is analytic in $\lambda$ for $\lambda\in\mathbb{C}\setminus\left(\left[-\eta_2, \eta_2\right]\cup\overline{\mathcal{C}_1}\cup\overline{\mathcal{C}_2}\right)$ and normalizes to
$
\begin{pmatrix}
1& 1
\end{pmatrix}
$
as  $\lambda\to\infty$,
where $\mathcal{C}_j=\mathcal{C}_{j, l}^+\cup\mathcal{C}_{j, r}^+\cup\mathcal{C}_{j, l}^-\cup\mathcal{C}_{j, r}^-$\, $(j=1,2)$.
For  $\lambda\in(-\eta_2, \eta_2)\cup\mathcal{C}_1\cup\mathcal{C}_2\setminus\left\{\pm\eta_1, \pm\alpha, \pm\eta_0\right\}$, $S$ admits continuous boundary values, which are related by the following jump conditions
\begin{gather}
S_+=S_-
\begin{cases}
\mathcal{U}^{tp}_{f}\left[\mathrm{i}r^{-1}\right], &\mathrm{for}\,\,\, \lambda\in\mathcal{C}_1,\\[0.5em]
\mathcal{L}^{tp}_{f}\left[-\mathrm{i}r^{-1}\right], &\mathrm{for}\,\,\, \lambda\in\mathcal{C}_2,\\[0.5em]
-\mathrm{i}\sigma_1, &\mathrm{for}\,\,\, \lambda\in\left(\alpha, \eta_0\right)\cup\left(\eta_0, \eta_2\right),\\[0.5em]
\mathrm{i}\sigma_1, &\mathrm{for}\,\,\, \lambda\in\left(-\eta_2, -\eta_0\right)\cup\left(-\eta_0, -\alpha\right),\\[0.5em]
f_-^{-\sigma_3}\mathrm{e}^{tp_-\sigma_3}\mathcal{L}\left[-\mathrm{i}r\right]\mathrm{e}^{-tp_+\sigma_3}f_+^{\sigma_3}, &\mathrm{for}\,\,\, \lambda\in\left(\eta_1, \alpha\right),\\[0.5em]
f_-^{-\sigma_3}\mathrm{e}^{tp_-\sigma_3}\mathcal{U}\left[\mathrm{i}r\right]\mathrm{e}^{-tp_+\sigma_3}f_+^{\sigma_3}, &\mathrm{for}\,\,\, \lambda\in\left(-\alpha, -\eta_1\right),\\[0.5em]
\mathrm{e}^{\left(\Omega t+\Delta\right)\sigma_3}, &\mathrm{for}\,\,\, \lambda\in\left(-\eta_1, \eta_1\right).
\end{cases}
\end{gather}
The local behaviors of $S$ near $\pm\eta_1$, $\pm\eta_2$, and $\pm\alpha$ are the same as those in the region $\xi_0<\xi<\eta_2^2$.
For the first generalized reflection coefficient $r_0$, the local behaviors near $\pm\eta_0$ are exhibited as follows:
if $\beta_0>0$, $S$ exhibits the following local behavior
\begin{gather}
S=
\begin{cases}
\mathcal{O}
\begin{pmatrix}
 \left|\lambda\mp\eta_0\right|^{\mp\beta_0/2}&\left|\lambda\mp\eta_0\right|^{\pm\beta_0/2}
\end{pmatrix}
, & \mathrm{as} \,\,\,\lambda\in\mathrm{D}_\mathrm{o}\to\pm\eta_0, \\[0.5em]
\mathcal{O}
\begin{pmatrix}
 \left|\lambda\mp\eta_0\right|^{-\beta_0/2}&\left|\lambda\mp\eta_0\right|^{-\beta_0/2}
\end{pmatrix}
, & \mathrm{as} \,\,\,\lambda\in\mathrm{D}_1\cup \mathrm{D}_2\to\pm\eta_0,
\end{cases}
\end{gather}
and if $\beta_0\in\left(-1, 0\right)$, the local behavior is formulated as
\begin{gather}
S=
\mathcal{O}
\begin{pmatrix}
 \left|\lambda\mp\eta_0\right|^{\beta_0/2}&\left|\lambda\mp\eta_0\right|^{\beta_0/2}
\end{pmatrix}
, \,\,\,\mathrm{as} \,\,\,\lambda\in\mathrm{D}_1\cup \mathrm{D}_2\cup \mathrm{D}_\mathrm{o}\to\pm\eta_0.
\end{gather}
For the second generalized coefficient $r_c$ near $\pm\eta_0$, $S$ exhibits the following local behavior
\begin{gather}
S=
\begin{cases}
\mathcal{O}
\begin{pmatrix}
\log\left|\lambda-\eta_0\right| &1
\end{pmatrix}
, & \mathrm{as} \,\,\,\lambda\in\mathrm{D}_\mathrm{o}\to\eta_0,\\[0.5em]
\mathcal{O}
\begin{pmatrix}
1 & \log\left|\lambda+\eta_0\right|
\end{pmatrix}
, & \mathrm{as} \,\,\,\lambda\in\mathrm{D}_\mathrm{o}\to-\eta_0,\\[0.5em]
\mathcal{O}
\begin{pmatrix}
\log\left|\lambda\mp\eta_0\right|  & \log\left|\lambda\mp\eta_0\right|
\end{pmatrix}
, & \mathrm{as} \,\,\,\lambda\in\mathrm{D}_1\cup \mathrm{D}_2\to\pm\eta_0.
\end{cases}
\end{gather}

\subsection{Local parametrix near the  singularity $\eta_0$ for the first generalized reflection coefficient  $r_0$}

For convenience, denote the domains  $\mathrm{D}_{\mathrm{o}, 1, l}^\pm$ and $\mathrm{D}_{\mathrm{o}, 1, r}^\pm$ as $\mathrm{D}_{\mathrm{o}, 1, l}^\pm=\mathrm{D}_\mathrm{o}^\pm\cap \left\{\lambda\in\mathbb{C}\mid\Re(\lambda)<\eta_0\right\}$ and
$\mathrm{D}_{\mathrm{o}, 1, r}^\pm=\mathrm{D}_\mathrm{o}^\pm\cap \left\{\lambda\in\mathbb{C}\mid\Re(\lambda)>\eta_0\right\}$.
For the first generalized reflection coefficient  $r_0$,  the local parametrix in the neighbourhood of $\lambda=\eta_0$ is constructed as follows:
\begin{itemize}

\item{} For $\lambda\in\left(\mathrm{D}_{1, r}^+\cup\mathrm{D}_{\mathrm{o}, 1, r}^+\right)\cap B(\eta_0)$, the local parametrix, denoted as $P^{\eta_0}$, is expressed as
\begin{gather}\label{parametrix-eta_0-1}
P^{\eta_0}=P^\infty\left(\frac{\mathrm{e}^{\pi\mathrm{i}/4+tp_+(\eta_0)}}{fd_r}\right)^{\sigma_3}\sigma_1\mathrm{e}^{\beta_0\pi\mathrm{i}\sigma_3/4}\left(-\mathrm{i}\sigma_2\right)M^{\mathrm{mb}}(\zeta_{\eta_0}, \beta_0)\mathrm{e}^{\zeta_{\eta_0}\sigma_3}\sigma_1\left(\frac{\mathrm{e}^{\pi\mathrm{i}/4+tp_+(\eta_0)}}{fd_r}\right)^{-\sigma_3},
\end{gather}
with $\zeta_{\eta_0}=t\left(p-p_+\left(\eta_0\right)\right)$;

\item {} For $\lambda\in \left(\mathrm{D}_{1, l}^+\cup\mathrm{D}_{\mathrm{o}, 1, l}^+\right)\cap B(\eta_0)$, $P^{\eta_0}$ is formulated as
\begin{gather}\label{parametrix-eta_0-2}
P^{\eta_0}=P^\infty\left(\frac{\mathrm{e}^{\pi\mathrm{i}/4+tp_+(\eta_0)}}{fd_l}\right)^{\sigma_3}\sigma_1\mathrm{e}^{-\beta_0\pi\mathrm{i}\sigma_3/4}\left(-\mathrm{i}\sigma_2\right)M^{\mathrm{mb}}(\zeta_{\eta_0}, \beta_0)\mathrm{e}^{\zeta_{\eta_0}\sigma_3}\sigma_1\left(\frac{\mathrm{e}^{\pi\mathrm{i}/4+tp_+(\eta_0)}}{fd_l}\right)^{-\sigma_3},
\end{gather}
with $\zeta_{\eta_0}=t\left(p-p_+\left(\eta_0\right)\right)$;

\item {} For $\lambda\in\left(\mathrm{D}_{1, l}^-\cup\mathrm{D}_{\mathrm{o}, 1, l}^-\right)\cap B(\eta_0)$,  $P^{\eta_0}$ is written as
\begin{gather}\label{parametrix-eta_0-3}
P^{\eta_0}=P^\infty\left(\frac{\mathrm{e}^{\pi\mathrm{i}/4+tp_-(\eta_0)}}{fd_l}\right)^{\sigma_3}\sigma_1\mathrm{e}^{\beta_0\pi\mathrm{i}\sigma_3/4}M^{\mathrm{mb}}(\zeta_{\eta_0}, \beta_0)\mathrm{e}^{-\zeta_{\eta_0}\sigma_3}\sigma_1\left(\frac{\mathrm{e}^{\pi\mathrm{i}/4+tp_-(\eta_0)}}{fd_l}\right)^{-\sigma_3},
\end{gather}
with $\zeta_{\eta_0}=-t\left(p-p_-\left(\eta_0\right)\right)$;

\item {} For $\lambda\in \left(\mathrm{D}_{1, r}^-\cup\mathrm{D}_{\mathrm{o}, 1, r}^-\right)\cap B(\eta_0)$, $P^{\eta_0}$ is expressed as
\begin{gather}\label{parametrix-eta_0-4}
P^{\eta_0}=P^\infty\left(\frac{\mathrm{e}^{\pi\mathrm{i}/4+tp_-(\eta_0)}}{fd_r}\right)^{\sigma_3}\sigma_1\mathrm{e}^{-\beta_0\pi\mathrm{i}\sigma_3/4}M^{\mathrm{mb}}(\zeta_{\eta_0}, \beta_0)\mathrm{e}^{-\zeta_{\eta_0}\sigma_3}\sigma_1\left(\frac{\mathrm{e}^{\pi\mathrm{i}/4+tp_-(\eta_0)}}{fd_r}\right)^{-\sigma_3},
\end{gather}
with $\zeta_{\eta_0}=-t\left(p-p_-\left(\eta_0\right)\right)$,
\end{itemize}
where the functions $d_l$ and $d_r$ are defined as follows:
\bee
\begin{array}{l}
d_l=(\lambda-\eta_1)^{\beta_1/2}(\eta_2-\lambda)^{\beta_2/2}(\lambda-\eta_0)^{\beta_0/2}\gamma(\lambda)^{1/2}, \vspace{0.15in}\\
d_r=
(\lambda-\eta_1)^{\beta_1/2}(\eta_2-\lambda)^{\beta_2/2}(\eta_0-\lambda)^{\beta_0/2}\gamma(\lambda)^{1/2},
\end{array}
\ene
and the matrix $M^{\mathrm{mb}}(\zeta_{\eta_0}, \beta_0)$ is constructed using the modified Bessel functions of the first and second kinds of index $\left(\beta_0\pm 1\right)\left/\right.2$.
It is important to note that the matrix $M^{\mathrm{mb}}$ is entirely different from $M^{\mathrm{mB}}$.
Previously, a modified Bessel parametrix analogous to $M^{\mathrm{mb}}$ was proposed in \cite{67} to study strong asymptotics of orthogonal polynomials with respect to a generalized Jacobi weight.
Ib this paper, $M^{\mathrm{mb}}(\zeta_{\eta_0}, \beta_0)$ is formulated as follows:
for $\lambda\in\mathrm{D}_{1, r}^+\cap B(\eta_0)$, the matrix $M^{\mathrm{mb}}(\zeta_{\eta_0}, \beta_0)$ is expressed as
\begin{gather}
M^{\mathrm{mb}}(\zeta_{\eta_0}, \beta_0)=C_1
\begin{pmatrix}
-G^+\left(\zeta_{\eta_0}\right) & G^+\left(\mathrm{e}^{-\pi\mathrm{i}}\zeta_{\eta_0}\right) \\[1em]
-G^-\left(\zeta_{\eta_0}\right) & G^-\left(\mathrm{e}^{-\pi\mathrm{i}}\zeta_{\eta_0}\right)
\end{pmatrix}
\mathrm{e}^{-\beta_0\pi\mathrm{i}\sigma_3/4};
\end{gather}
for $\lambda\in\mathrm{D}_{\mathrm{o}, 1, r}^+\cap B(\eta_0)$, the matrix $M^{\mathrm{mb}}(\zeta_{\eta_0}, \beta_0)$ is expressed as
\begin{gather}
M^{\mathrm{mb}}(\zeta_{\eta_0}, \beta_0)=C_1
\begin{pmatrix}
-H^+\left(\mathrm{e}^{-\pi\mathrm{i}}\zeta_{\eta_0}\right) & G^+\left(\mathrm{e}^{-\pi\mathrm{i}}\zeta_{\eta_0}\right) \\[1em]
-H^-\left(\mathrm{e}^{-\pi\mathrm{i}}\zeta_{\eta_0}\right) & -G^-\left(\mathrm{e}^{-\pi\mathrm{i}}\zeta_{\eta_0}\right)
\end{pmatrix}
\mathrm{e}^{-\beta_0\pi\mathrm{i}\sigma_3/4};
\end{gather}
for $\lambda\in\mathrm{D}_{\mathrm{o}, 1, l}^+\cap B(\eta_0)$, the matrix $M^{\mathrm{mb}}(\zeta_{\eta_0}, \beta_0)$ is expressed as
\begin{gather}
M^{\mathrm{mb}}(\zeta_{\eta_0}, \beta_0)=C_1
\begin{pmatrix}
-H^+\left(\mathrm{e}^{-\pi\mathrm{i}}\zeta_{\eta_0}\right) & G^+\left(\mathrm{e}^{-\pi\mathrm{i}}\zeta_{\eta_0}\right) \\[1em]
-H^-\left(\mathrm{e}^{-\pi\mathrm{i}}\zeta_{\eta_0}\right) & -G^-\left(\mathrm{e}^{-\pi\mathrm{i}}\zeta_{\eta_0}\right)
\end{pmatrix}
\mathrm{e}^{\beta_0\pi\mathrm{i}\sigma_3/4};
\end{gather}
for $\lambda\in\mathrm{D}_{1, l}^+\cap B(\eta_0)$, the matrix $M^{\mathrm{mb}}(\zeta_{\eta_0}, \beta_0)$ is expressed as
\begin{gather}
M^{\mathrm{mb}}(\zeta_{\eta_0}, \beta_0)=C_1
\begin{pmatrix}
-G^+\left(\mathrm{e}^{-2\pi\mathrm{i}}\zeta_{\eta_0}\right) & G^+\left(\mathrm{e}^{-\pi\mathrm{i}}\zeta_{\eta_0}\right) \\[0.5em]
-G^-\left(\mathrm{e}^{-2\pi\mathrm{i}}\zeta_{\eta_0}\right) & -G^-\left(\mathrm{e}^{-\pi\mathrm{i}}\zeta_{\eta_0}\right)
\end{pmatrix}
\mathrm{e}^{\beta_0\pi\mathrm{i}\sigma_3/4};
\end{gather}
for $\lambda\in\mathrm{D}_{1, l}^-\cap B(\eta_0)$, the matrix $M^{\mathrm{mb}}(\zeta_{\eta_0}, \beta_0)$ is expressed as
\begin{gather}
M^{\mathrm{mb}}(\zeta_{\eta_0}, \beta_0)=C_1
\begin{pmatrix}
G^+\left(\mathrm{e}^{\pi\mathrm{i}}\zeta_{\eta_0}\right) & G^+\left(\zeta_{\eta_0}\right) \\[0.5em]
-G^-\left(\mathrm{e}^{\pi\mathrm{i}}\zeta_{\eta_0}\right) & G^-\left(\zeta_{\eta_0}\right)
\end{pmatrix}
\mathrm{e}^{-\beta_0\pi\mathrm{i}\sigma_3/4};
\end{gather}
for $\lambda\in\mathrm{D}_{\mathrm{o}, 1, l}^-\cap B(\eta_0)$, the matrix $M^{\mathrm{mb}}(\zeta_{\eta_0}, \beta_0)$ is expressed as
\begin{gather}
M^{\mathrm{mb}}(\zeta_{\eta_0}, \beta_0)=C_1
\begin{pmatrix}
H^+\left(\zeta_{\eta_0}\right) & G^+\left(\zeta_{\eta_0}\right) \\[0.5em]
-H^-\left(\zeta_{\eta_0}\right) & G^-\left(\zeta_{\eta_0}\right)
\end{pmatrix}
\mathrm{e}^{-\beta_0\pi\mathrm{i}\sigma_3/4};
\end{gather}
for $\lambda\in\mathrm{D}_{\mathrm{o}, 1, r}^-\cap B(\eta_0)$, the matrix $M^{\mathrm{mb}}(\zeta_{\eta_0}, \beta_0)$ is expressed as
\begin{gather}
M^{\mathrm{mb}}(\zeta_{\eta_0}, \beta_0)=C_1
\begin{pmatrix}
H^+\left(\zeta_{\eta_0}\right) & G^+\left(\zeta_{\eta_0}\right) \\[0.5em]
-H^-\left(\zeta_{\eta_0}\right) & G^-\left(\zeta_{\eta_0}\right)
\end{pmatrix}
\mathrm{e}^{\beta_0\pi\mathrm{i}\sigma_3/4};
\end{gather}
for $\lambda\in\mathrm{D}_{1, r}^- \cap B(\eta_0)$, the matrix $M^{\mathrm{mb}}(\zeta_{\eta_0}, \beta_0)$ is expressed as
\begin{gather}
M^{\mathrm{mb}}(\zeta_{\eta_0}, \beta_0)=C_1
\begin{pmatrix}
G^+\left(\mathrm{e}^{-\pi\mathrm{i}}\zeta_{\eta_0}\right) & G^+\left(\zeta_{\eta_0}\right) \\[0.5em]
-G^-\left(\mathrm{e}^{-\pi\mathrm{i}}\zeta_{\eta_0}\right) & G^-\left(\zeta_{\eta_0}\right)
\end{pmatrix}
\mathrm{e}^{\beta_0\pi\mathrm{i}\sigma_3/4},
\end{gather}
where $G^\pm: \zeta_{\eta_0}\mapsto G^\pm(\zeta_{\eta_0})$ is a function with $G^\pm\left(\zeta_{\eta_0}\right)=\sqrt{\zeta_{\eta_0}/\pi}\,K_{\left(\beta_0\pm1\right)/2}\left(\zeta_{\eta_0}\right)$
and $H^\pm: \zeta_{\eta_0}\mapsto H^\pm(\zeta_{\eta_0})$ is a function with
$H^\pm\left(\zeta_{\eta_0}\right)=\sqrt{\pi\zeta_{\eta_0}}\,I_{\left(\beta_0\pm1\right)/2}\left(\zeta_{\eta_0}\right)$,
where the argument of $\zeta_{\eta_0}$ is subject to $\left(-\pi/2,\,3\pi/2\right)$.
$M^{\mathrm{mb}}\left(\zeta_{\eta_0}, \beta_0\right)$ solves a $2\times 2$ Riemann-Hilbert problem with the following properties, as depicted in Figure \ref{BesselCH}(Left).
$M^{\mathrm{mb}}\left(\zeta_{\eta_0}, \beta_0\right)$ is analytic in $\zeta_{\eta_0}$ for $\zeta_{\eta_0}\in\mathbb{C}\setminus(\cup_{j=1}^8\Sigma_j)$ and normalizes to
\begin{gather}
M^{\mathrm{mb}}\left(\zeta_{\eta_0}, \beta_0\right)=
\begin{cases}
\left(\mathbb{I}_2+\mathcal{O}\left(\zeta_{\eta_0}^{-1}\right)\right)\mathrm{i}\sigma_2\,\mathrm{e}^{-\beta_0\pi\mathrm{i}\sigma_3/4}\mathrm{e}^{-\zeta_{\eta_0}\sigma_3}, &\mathrm{as}\,\,\, \zeta_{\eta_0}\in\mathrm{D}^\zeta_1\cup\mathrm{D}^\zeta_2\to\infty, \\[0.5em]
\left(\mathbb{I}_2+\mathcal{O}\left(\zeta_{\eta_0}^{-1}\right)\right)\mathrm{i}\sigma_2\,\mathrm{e}^{\beta_0\pi\mathrm{i}\sigma_3/4}\mathrm{e}^{-\zeta_{\eta_0}\sigma_3}, & \mathrm{as}\,\,\, \zeta_{\eta_0}\in\mathrm{D}^\zeta_3\cup\mathrm{D}^\zeta_4\to\infty, \\[0.5em]
\left(\mathbb{I}_2+\mathcal{O}\left(\zeta_{\eta_0}^{-1}\right)\right)\mathrm{e}^{-\beta_0\pi\mathrm{i}\sigma_3/4}\mathrm{e}^{\zeta_{\eta_0}\sigma_3}, &\mathrm{as}\,\,\, \zeta_{\eta_0}\in\mathrm{D}^\zeta_5\cup\mathrm{D}^\zeta_6\to\infty, \\[0.5em]
\left(\mathbb{I}_2+\mathcal{O}\left(\zeta_{\eta_0}^{-1}\right)\right)\mathrm{e}^{\beta_0\pi\mathrm{i}\sigma_3/4}\mathrm{e}^{\zeta_{\eta_0}\sigma_3}, & \mathrm{as}\,\,\, \zeta_{\eta_0}\in\mathrm{D}^\zeta_7\cup\mathrm{D}^\zeta_8\to\infty.
\end{cases}
\end{gather}
For $\zeta_{\eta_0}\in\cup_{j=1}^8\Sigma_j^0$, $M^{\mathrm{mb}}\left(\zeta_{\eta_0}, \beta_0\right)$ admits continuous boundary values, which are related by the following jump conditions
\begin{gather}
M^{\mathrm{mb}}_+\left(\zeta_{\eta_0}, \beta_0\right)=M^{\mathrm{mb}}_-\left(\zeta_{\eta_0}, \beta_0\right)
\begin{cases}
\mathcal{L}\left[\mathrm{e}^{-\beta_0 \pi \mathrm{i}}\right], &\mathrm{for}\,\,\, \zeta_{\eta_0}\in\Sigma_1^0\cup\Sigma_5^0, \\[0.5em]
\mathrm{e}^{\beta_0\pi\mathrm{i}\sigma_3/2}, &\mathrm{for}\,\,\, \zeta_{\eta_0}\in\Sigma_2^0\cup\Sigma_6^0, \\[0.5em]
\mathcal{L}\left[\mathrm{e}^{\beta_0 \pi \mathrm{i}}\right], &\mathrm{for}\,\,\, \zeta_{\eta_0}\in\Sigma_3^0\cup\Sigma_7^0, \\[0.5em]
\mathrm{i}\sigma_2, &\mathrm{for}\,\,\, \zeta_{\eta_0}\in\Sigma_4^0\cup\Sigma_8^0,
\end{cases}
\end{gather}
where $\Sigma_j^0=\Sigma_j\setminus \{0\}, \, j=1, 2, \cdots, 8$. Near the origin for $\beta_0>-1$ and $\beta_0\ne 0$, the matrix $M^{\mathrm{mb}}\left(\zeta_{\eta_0}, \beta_0\right)$ exhibits the following local behaviors
\begin{gather}
M^{\mathrm{mb}}\left(\zeta_{\eta_0}, \beta_0\right)=
\begin{cases}
\mathcal{O}
\begin{pmatrix}
\left|\zeta_{\eta_0}\right|^{\beta_0/2} & \left|\zeta_{\eta_0}\right|^{-\left|\beta_0\right|/2}  \\[0.5em]
\left|\zeta_{\eta_0}\right|^{\beta_0/2} & \left|\zeta_{\eta_0}\right|^{-\left|\beta_0\right|/2}
\end{pmatrix},
& \mathrm{as}\,\,\, \zeta_{\eta_0}\in \mathrm{D}^\zeta_2\cup\mathrm{D}^\zeta_3\cup\mathrm{D}^\zeta_6\cup\mathrm{D}^\zeta_7\to 0, \\[2em]
\mathcal{O}
\begin{pmatrix}
\left|\zeta_{\eta_0}\right|^{-\left|\beta_0\right|/2} & \left|\zeta_{\eta_0}\right|^{-\left|\beta_0\right|/2}  \\[0.5em]
\left|\zeta_{\eta_0}\right|^{-\left|\beta_0\right|/2} & \left|\zeta_{\eta_0}\right|^{-\left|\beta_0\right|/2}
\end{pmatrix},
& \mathrm{as}\,\,\, \zeta_{\eta_0}\in \mathrm{D}^\zeta_1\cup\mathrm{D}^\zeta_4\cup\mathrm{D}^\zeta_5\cup\mathrm{D}^\zeta_8\to 0.
\end{cases}
\end{gather}

\begin{figure}[!t]
\centering
\includegraphics[scale=0.26]{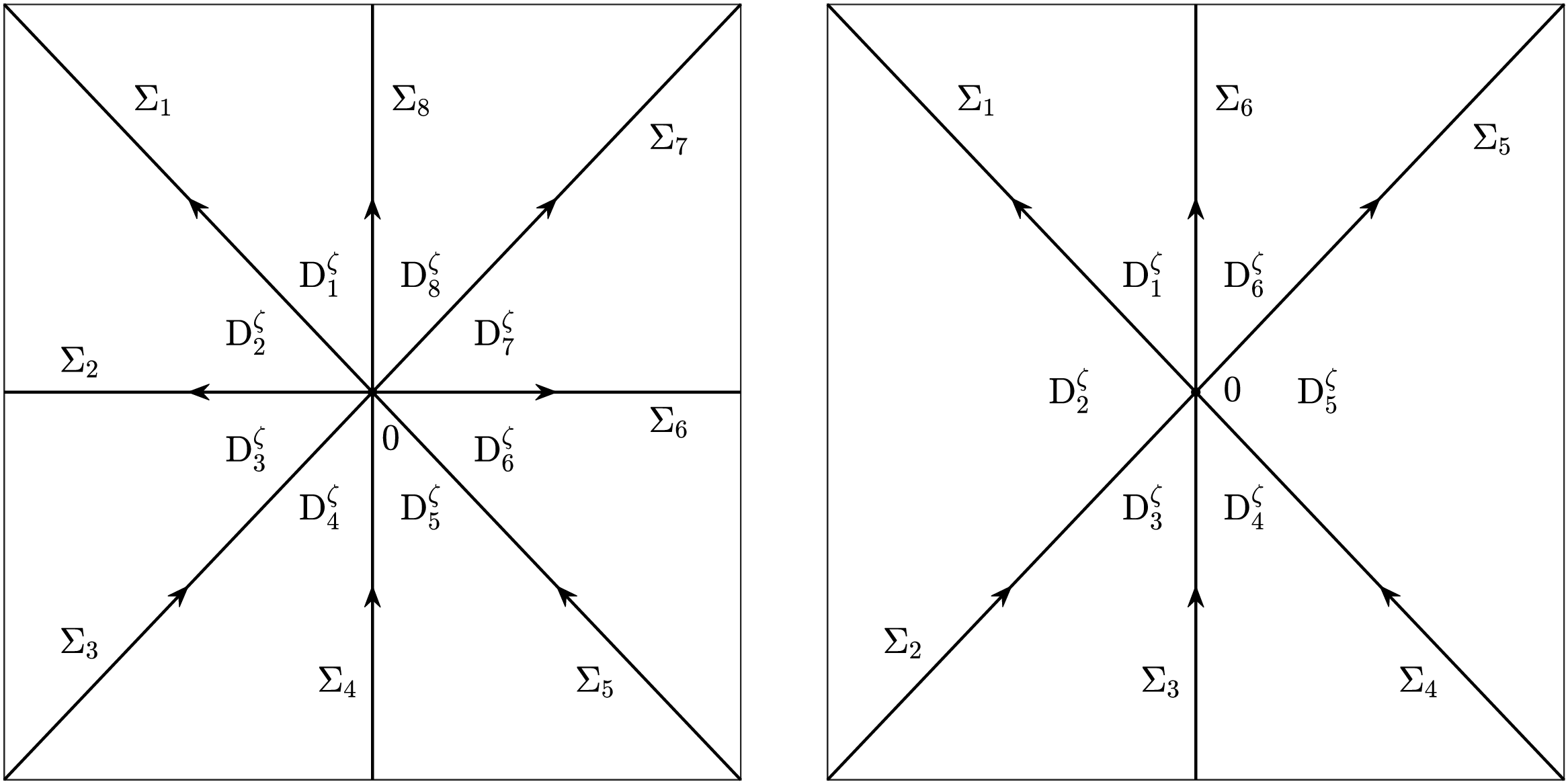}
\caption{Left: Jump contour for modified Bessel parametrix $M^{\mathrm{mb}}$; Right: Jump contour for Confluent Hypergeometric parametrix $M^{\mathrm{CH}}$.}
\label{BesselCH}
\end{figure}

It follows from expressions \eqref{parametrix-eta_0-1}, \eqref{parametrix-eta_0-2}, \eqref{parametrix-eta_0-3}, and \eqref{parametrix-eta_0-4} that $P^{\eta_0}$ solves a Riemann-Hilbert problem in the neighborhood of $\lambda=\eta_0$.
The function $P^{\eta_0}$ is analytic in $\lambda$ for $\lambda\in \left(\mathrm{D}_1\cup \mathrm{D}_\mathrm{o}\right)\cap B(\eta_0)$, and as $t\to +\infty$, it normalizes to
\begin{gather}\label{small-norm-eta_0}
P^{\eta_0}\left(P^\infty\right)^{-1}=\mathbb{I}_2+\mathcal{O}\left(\frac{1}{t}\right).
\end{gather}
For $\lambda\in \left(\mathcal{C}_1\cup \left(\alpha, \eta_2\right)\right)\cap B(\eta_0)$, $P^{\eta_0}$ admits continuous boundary values, denoted as $P^{\eta_0}_+$ and $P^{\eta_0}_-$, related by the following jump relations
\begin{gather}\label{jump-eta_0}
P^{\eta_0}_+=P^{\eta_0}_-
\begin{cases}
\mathcal{U}^{tp}_{f}\left[\mathrm{i}r^{-1}\right], &\mathrm{for}\,\,\, \lambda\in\mathcal{C}_1\cap B(\eta_0),\\[0.5em]
-\mathrm{i}\sigma_1, &\mathrm{for}\,\,\, \lambda\in\left(\left(\alpha, \eta_0\right)\cup\left(\eta_0, \eta_2\right)\right)\cap B(\eta_0).
\end{cases}
\end{gather}
For the first generalized reflection coefficient, $r=r_0$, the matrix $P^{\eta_0}$ closely matches the behaviors of $S$ near $\lambda=\eta_0$ and exhibits the following local behaviors:
if $\beta_0>0$, $P^{\eta_0}$ exhibits locally as
\begin{gather}
P^{\eta_0}=
\begin{cases}
\mathcal{O}
\begin{pmatrix}
 \left|\lambda-\eta_0\right|^{-\beta_0/2}&\left|\lambda-\eta_0\right|^{\beta_0/2}  \\[0.5em]
  \left|\lambda-\eta_0\right|^{-\beta_0/2}&\left|\lambda-\eta_0\right|^{\beta_0/2}
\end{pmatrix}
, & \mathrm{as} \,\,\,\lambda\in\mathrm{D}_\mathrm{o}\to\eta_0, \\[2em]
\mathcal{O}
\begin{pmatrix}
 \left|\lambda-\eta_0\right|^{-\beta_0/2}&\left|\lambda-\eta_0\right|^{-\beta_0/2}\\[0.5em]
  \left|\lambda-\eta_0\right|^{-\beta_0/2}&\left|\lambda-\eta_0\right|^{-\beta_0/2}
\end{pmatrix}
, & \mathrm{as} \,\,\,\lambda\in\mathrm{D}_1 \to\eta_0,
\end{cases}
\end{gather}
and if $\beta_0\in\left(-1, 0\right)$, the local behavior is given by
\begin{gather}
P^{\eta_0}=
\mathcal{O}
\begin{pmatrix}
 \left|\lambda-\eta_0\right|^{\beta_0/2}&\left|\lambda-\eta_0\right|^{\beta_0/2} \\[0.5em]
 \left|\lambda-\eta_0\right|^{\beta_0/2}&\left|\lambda-\eta_0\right|^{\beta_0/2}
\end{pmatrix}
, \,\,\,\mathrm{as} \,\,\,\lambda\in\mathrm{D}_1\cup \mathrm{D}_\mathrm{o}\to\eta_0.
\end{gather}

\subsection{Local parametrix near $\eta_0$ for the second generalized reflection coefficient  $r_c$}

For the second generalized reflection coefficient  $r_c$,  the local parametrix in the neighbourhood of $\lambda=\eta_0$ is constructed as follows:
\begin{itemize}

\item {} For $\lambda\in \left(\mathrm{D}_1^+\cup \mathrm{D}_\mathrm{o, 1}^+\right)\cap B(\eta_0)$, the local parametrix, denoted as $P^{\eta_0}$, is formulated as
\begin{gather}\label{parametrix-eta_0-a}
\begin{array}{rl}
P^{\eta_0}=& P^\infty\left(\dfrac{\mathrm{e}^{\pi\mathrm{i}/4+tp_+(\eta_0)}}{fd}\right)^{\sigma_3}
\sigma_1\left(\zeta_{\eta_0}^{\kappa_0\sigma_3}\mathrm{i}\sigma_2\mathrm{e}^{\kappa_0\pi\mathrm{i}\sigma_3}\right)^{-1} \\ [1em]
&\times M^{\mathrm{CH}}(\zeta_{\eta_0}, \kappa_0)\mathrm{e}^{\zeta_{\eta_0}\sigma_3/2}\sigma_1\left(\dfrac{\mathrm{e}^{\pi\mathrm{i}/4+tp_+(\eta_0)}}{fd}\right)^{-\sigma_3}
\end{array}
\end{gather}
with $\zeta_{\eta_0}=2t\left(p-p_+\left(\eta_0\right)\right)$;

\item {} For $\lambda\in \left(\mathrm{D}_1^-\cup \mathrm{D}_\mathrm{o, 1}^-\right)\cap B(\eta_0)$, the local parametrix $P^{\eta_0}$ is expressed as
\begin{gather}\label{parametrix-eta_0-b}
P^{\eta_0}=P^\infty\left(\frac{\mathrm{e}^{\pi\mathrm{i}/4+tp_-(\eta_0)}}{fd}\right)^{\sigma_3}\sigma_1\mathrm{e}^{-\kappa_0\pi\mathrm{i}\sigma_3}M^{\mathrm{CH}}(\zeta_{\eta_0},  \kappa_0)\mathrm{e}^{-\zeta_{\eta_0}\sigma_3/2}\sigma_1\left(\frac{\mathrm{e}^{\pi\mathrm{i}/4+tp_-(\eta_0)}}{fd}\right)^{-\sigma_3},
\end{gather}
with $\zeta_{\eta_0}=-2t\left(p-p_-\left(\eta_0\right)\right)$,
\end{itemize}
where $\kappa_0=\left(\mathrm{i}/\pi\right)\log c$ and
$d=\left(\lambda-\eta_1\right)^{\kappa_1/2}\left(\eta_2-\lambda\right)^{\kappa_2/2}c^{1/2}\gamma(\lambda)^{1/2}$,
 $M^{\mathrm{CH}}(\zeta_{\eta_0}, \kappa_0)$ is constructed by the confluent hypergeometric functions $M(\zeta_{\eta_0}, \kappa_0)$ and $U(\zeta_{\eta_0}, \kappa_0)$, which are two standard solutions of a special Kummer's equation:
\bee
 \zeta_{\eta_0} \,y''(\zeta_{\eta_0})+(1-\zeta_{\eta_0})y'(\zeta_{\eta_0})-\kappa_0 y(\zeta_{\eta_0})=0,
\ene
where $M(\zeta_{\eta_0}, \kappa_0)$ is a entire function, written as
$M(\zeta_{\eta_0}, \kappa_0)=\sum_{n=1}^\infty \Gamma\left(\kappa_0+n\right)\zeta_{\eta_0}^n\left/\right.\Gamma\left(\kappa_0\right) n!,
$
and $U(\zeta_{\eta_0}, \kappa_0)$ has a branch point at $\zeta_{\eta_0}=0$, uniquely determined by the asymptotic property:
$
U(\zeta_{\eta_0}, \kappa_0)\sim \zeta_{\eta_0}^{-\kappa_0},
$
as $\zeta_{\eta_0}\to\infty$ for $ \mathrm{arg}\, \zeta_{\eta_0}\in (-3\pi /2, 3\pi/2)$.
The confluent hypergeometric parametrix was constructed in \cite{68} in the study of asymptotic behaviors of Hanker determinants and  orthogonal polynomials with respect to a Gaussian weight with a jump.
Through a slight modification, the confluent hypergeometric parametrix, denoted as $M^{\mathrm{CH}}(\zeta_{\eta_0}, \kappa_0)$ used in this paper, is formulated as follows:
for  $\lambda\in\mathrm{D}_{1, r}^+\cap B(\eta_0)$, the matrix $M^{\mathrm{CH}}(\zeta_{\eta_0}, \kappa_0)$ is expressed as
\begin{gather}
M^{\mathrm{CH}}(\zeta_{\eta_0}, \kappa_0)=
\begin{pmatrix}
-\dfrac{\Gamma\left(1+\kappa_0\right)}{\Gamma\left(-\kappa_0\right)}\,\mathrm{e}^{\kappa_0\pi\mathrm{i}}U\left(1+\kappa_0, \zeta_{\eta_0}\right) & U\left(-\kappa_0, \mathrm{e}^{-\pi\mathrm{i}}\zeta_{\eta_0}\right) \\[1em]
-\mathrm{e}^{\kappa_0\pi\mathrm{i}}U\left(\kappa_0, \zeta_{\eta_0}\right)& \dfrac{\Gamma\left(1-\kappa_0\right)}{\Gamma\left(\kappa_0\right)}U\left(1-\kappa_0,\mathrm{e}^{-\pi\mathrm{i}}\zeta_{\eta_0}\right)
\end{pmatrix}
\mathrm{e}^{-\zeta_{\eta_0}\sigma_3/2};
\end{gather}
for  $\lambda\in\mathrm{D}_\mathrm{o}^+\cap B(\eta_0)$, the matrix $M^{\mathrm{CH}}(\zeta_{\eta_0}, \kappa_0)$ is expressed as
\begin{gather}
M^{\mathrm{CH}}(\zeta_{\eta_0}, \kappa_0)=
\begin{pmatrix}
\Gamma\left(1+\kappa_0\right)M\left(-\kappa_0, \mathrm{e}^{-\pi\mathrm{i}}\zeta_{\eta_0}\right) & U\left(-\kappa_0, \mathrm{e}^{-\pi\mathrm{i}}\zeta_{\eta_0}\right) \\[1em]
-\Gamma\left(1-\kappa_0\right)M\left(1-\kappa_0, \mathrm{e}^{-\pi\mathrm{i}}\zeta_{\eta_0}\right)& \dfrac{\Gamma\left(1-\kappa_0\right)}{\Gamma\left(\kappa_0\right)}U\left(1-\kappa_0,\mathrm{e}^{-\pi\mathrm{i}}\zeta_{\eta_0}\right)
\end{pmatrix}
\mathrm{e}^{\zeta_{\eta_0}/2};
\end{gather}
for  $\lambda\in\mathrm{D}^+_{1, l}\cap B(\eta_0)$, the matrix $M^{\mathrm{CH}}(\zeta_{\eta_0}, \kappa_0)$ is expressed as
\begin{gather}
M^{\mathrm{CH}}(\zeta_{\eta_0}, \kappa_0)=
\begin{pmatrix}
-\dfrac{\Gamma\left(1+\kappa_0\right)}{\Gamma\left(-\kappa_0\right)}\,\mathrm{e}^{-\kappa_0\pi\mathrm{i}}U\left(1+\kappa_0, \mathrm{e}^{-2\pi\mathrm{i}}\zeta_{\eta_0}\right) & U\left(-\kappa_0, \mathrm{e}^{-\pi\mathrm{i}}\zeta_{\eta_0}\right) \\[1em]
-\mathrm{e}^{-\kappa_0\pi\mathrm{i}}U\left(\kappa_0, \mathrm{e}^{-2\pi\mathrm{i}}\zeta_{\eta_0}\right)& \dfrac{\Gamma\left(1-\kappa_0\right)}{\Gamma\left(\kappa_0\right)}U\left(1-\kappa_0,\mathrm{e}^{-\pi\mathrm{i}}\zeta_{\eta_0}\right)
\end{pmatrix}
\mathrm{e}^{-\zeta_{\eta_0}\sigma_3/2};
\end{gather}
for  $\lambda\in\mathrm{D}_{1, l}^-\cap B(\eta_0)$, the matrix $M^{\mathrm{CH}}(\zeta_{\eta_0}, \kappa_0)$ is expressed as
\begin{gather}
M^{\mathrm{CH}}(\zeta_{\eta_0}, \kappa_0)=
\begin{pmatrix}
\mathrm{e}^{-\kappa_0\pi\mathrm{i}}U\left(-\kappa_0, \mathrm{e}^{\pi\mathrm{i}}\zeta_{\eta_0}\right) & \dfrac{\Gamma\left(1+\kappa_0\right)}{\Gamma\left(-\kappa_0\right)}U\left(1+\kappa_0, \zeta_{\eta_0}\right) \\[1em]
\dfrac{\Gamma\left(1-\kappa_0\right)}{\Gamma\left(\kappa_0\right)}\mathrm{e}^{-\kappa_0\pi\mathrm{i}}U\left(1-\kappa_0, \mathrm{e}^{\pi\mathrm{i}}\zeta_{\eta_0}\right) & U\left(\kappa_0, \zeta_{\eta_0}\right)
\end{pmatrix}
\mathrm{e}^{\zeta_{\eta_0}\sigma_3/2};
\end{gather}
for  $\lambda\in\mathrm{D}_\mathrm{o}^-\cap B(\eta_0)$, the matrix $M^{\mathrm{CH}}(\zeta_{\eta_0}, \kappa_0)$ is expressed as
\begin{gather}
M^{\mathrm{CH}}(\zeta_{\eta_0}, \kappa_0)=
\begin{pmatrix}
\Gamma\left(1+\kappa_0\right)M\left(1+\kappa_0, \zeta_{\eta_0}\right) & \dfrac{\Gamma\left(1+\kappa_0\right)}{\Gamma\left(-\kappa_0\right)}U\left(1+\kappa_0, \zeta_{\eta_0}\right) \\[1em]
-\Gamma\left(1-\kappa_0\right)M\left(\kappa_0, \zeta_{\eta_0}\right) & U\left(\kappa_0, \zeta_{\eta_0}\right)
\end{pmatrix}
\mathrm{e}^{-\zeta_{\eta_0}/2},
\end{gather}
for  $\lambda\in\mathrm{D}_{1, r}^-\cap B(\eta_0)$, the matrix $M^{\mathrm{CH}}(\zeta_{\eta_0}, \kappa_0)$ is expressed as
\begin{gather}
M^{\mathrm{CH}}(\zeta_{\eta_0}, \kappa_0)=
\begin{pmatrix}
\mathrm{e}^{\kappa_0\pi\mathrm{i}}U\left(-\kappa_0, \mathrm{e}^{-\pi\mathrm{i}}\zeta_{\eta_0}\right) & \dfrac{\Gamma\left(1+\kappa_0\right)}{\Gamma\left(-\kappa_0\right)}U\left(1+\kappa_0, \zeta_{\eta_0}\right) \\[1em]
\dfrac{\Gamma\left(1-\kappa_0\right)}{\Gamma\left(\kappa_0\right)}\mathrm{e}^{\kappa_0\pi\mathrm{i}}U\left(1-\kappa_0, \mathrm{e}^{-\pi\mathrm{i}}\zeta_{\eta_0}\right) & U\left(\kappa_0, \zeta_{\eta_0}\right)
\end{pmatrix}
\mathrm{e}^{\zeta_{\eta_0}\sigma_3/2}.
\end{gather}
$M^{\mathrm{CH}}\left(\zeta_{\eta_0}, \kappa_0\right)$ solves a Riemann-Hilbert problem with the following properties, as depicted in Figure \ref{BesselCH}(Right).
$M^{\mathrm{CH}}\left(\zeta_{\eta_0}, \kappa_0\right)$ is analytic in $\zeta_{\eta_0}$ for $\zeta_{\eta_0}\in\mathbb{C}\setminus\left(\cup_{j=1}^6\Sigma_j\right)$ and normalizes to
\begin{gather}
M^{\mathrm{CH}}\left(\zeta_{\eta_0}, \kappa_0\right)=
\begin{cases}
\left(\mathbb{I}_2+\mathcal{O}\left(\zeta_{\eta_0}^{-1}\right)\right)\zeta_{\eta_0}^{\kappa_0\sigma_3}\mathrm{e}^{\zeta_{\eta_0}\sigma_3/2}, &\mathrm{as}\,\,\, \zeta_{\eta_0}\in\mathrm{D}^\zeta_4\cup\mathrm{D}^\zeta_5\cup\mathrm{D}^\zeta_6\to\infty, \\[1em]
\left(\mathbb{I}_2+\mathcal{O}\left(\zeta_{\eta_0}^{-1}\right)\right)\mathrm{i}\sigma_2\mathrm{e}^{\kappa_0\pi\mathrm{i}\sigma_3}\zeta_{\eta_0}^{-\kappa_0\sigma_3}\mathrm{e}^{-\zeta_{\eta_0}\sigma_3}, & \mathrm{as}\,\,\, \zeta_{\eta_0}\in\mathrm{D}^\zeta_1\cup\mathrm{D}^\zeta_2\cup\mathrm{D}^\zeta_3\to\infty.
\end{cases}
\end{gather}
For $\zeta_{\eta_0}\in\cup_{j=1}^6\Sigma_j^0$, $M^{\mathrm{CH}}\left(\zeta_{\eta_0}, \kappa_0\right)$ admits continuous boundary values  denoted by $M^{\mathrm{CH}}_+\left(\zeta_{\eta_0}, \kappa_0\right)$ and $M^{\mathrm{CH}}_-\left(\zeta_{\eta_0}, \kappa_0\right)$,  respectively. These values are related by the following jump conditions
\begin{gather}
M^{\mathrm{CH}}_+\left(\zeta_{\eta_0}, \kappa_0\right)=M^{\mathrm{CH}}_-\left(\zeta_{\eta_0}, \kappa_0\right)
\begin{cases}
\mathcal{L}\left[\mathrm{e}^{\kappa_0 \pi \mathrm{i}}\right], &\mathrm{for}\,\,\, \zeta_{\eta_0}\in\Sigma_1^0\cup\Sigma_5^0, \\[0.5em]
\mathcal{L}\left[\mathrm{e}^{-\kappa_0 \pi \mathrm{i}}\right], &\mathrm{for}\,\,\, \zeta_{\eta_0}\in\Sigma_2^0\cup\Sigma_4^0, \\[0.5em]
\mathrm{i}\sigma_2\,\mathrm{e}^{-\kappa_0\pi\mathrm{i}\sigma_3}, &\mathrm{for}\,\,\, \zeta_{\eta_0}\in\Sigma_3^0, \\[0.5em]
\mathrm{i}\sigma_2\,\mathrm{e}^{\kappa_0\pi\mathrm{i}\sigma_3}, &\mathrm{for}\,\,\, \zeta_{\eta_0}\in\Sigma_6^0,
\end{cases}
\end{gather}
 where $\Sigma_j^0=\Sigma_j\setminus \{0\}, \, j=1, 2, \cdots, 6$.
 $M^{\mathrm{CH}}\left(\zeta_{\eta_0}, \kappa_0\right)$ exhibits the following local behaviors near the origin
\begin{gather}
M^{\mathrm{CH}}\left(\zeta_{\eta_0}, \kappa_0\right)=
\begin{cases}
\mathcal{O}
\begin{pmatrix}
1 & \log\left|\zeta_{\eta_0}\right|  \\[0.5em]
1 & \log\left|\zeta_{\eta_0}\right|
\end{pmatrix},
& \mathrm{as}\,\,\, \zeta_{\eta_0}\in \mathrm{D}^\zeta_2\cup\mathrm{D}^\zeta_5\to 0, \\[2em]
\mathcal{O}
\begin{pmatrix}
\log\left|\zeta_{\eta_0}\right| & \log\left|\zeta_{\eta_0}\right|  \\[0.5em]
\log\left|\zeta_{\eta_0}\right| & \log\left|\zeta_{\eta_0}\right|
\end{pmatrix},
& \mathrm{as}\,\,\, \zeta_{\eta_0}\in \mathrm{D}^\zeta_1\cup\mathrm{D}^\zeta_3\cup\mathrm{D}^\zeta_4\cup\mathrm{D}^\zeta_6\to 0.
\end{cases}
\end{gather}
It follows from expressions \eqref{parametrix-eta_0-a} and \eqref{parametrix-eta_0-b} that $P^{\eta_0}$ solves a Riemann-Hilbert problem in the neighborhood of $\lambda=\eta_0$.
$P^{\eta_0}$ is analytic in $\lambda$ for $\lambda\in \left(\mathrm{D}_1\cup \mathrm{D}_\mathrm{o}\right)\cap B(\eta_0)$, and as $t\to +\infty$, it normalizes as \eqref{small-norm-eta_0}.
For $\lambda\in \left(\mathcal{C}_1\cup \left(\alpha, \eta_2\right)\right)\cap B(\eta_0)$, $P^{\eta_0}$ admits continuous boundary values, denoted as $P^{\eta_0}_+$ and $P^{\eta_0}_-$, which are related by the jump relations \eqref{jump-eta_0}.
For the second generalized coefficient $r_c$ near $\pm\eta_0$, the matrix $P^{\eta_0}$ exhibits the following local behavior
\begin{gather}
P^{\eta_0}=
\begin{cases}
\mathcal{O}
\begin{pmatrix}
\log\left|\lambda-\eta_0\right| &1  \\[0.5em]
\log\left|\lambda-\eta_0\right| &1
\end{pmatrix}
, & \mathrm{as} \,\,\,\lambda\in\mathrm{D}_\mathrm{o}\to\eta_0,\\[2em]
\mathcal{O}
\begin{pmatrix}
\log\left|\lambda-\eta_0\right|  & \log\left|\lambda-\eta_0\right| \\[0.5em]
\log\left|\lambda-\eta_0\right|  & \log\left|\lambda-\eta_0\right|
\end{pmatrix}
, & \mathrm{as} \,\,\,\lambda\in\mathrm{D}_1\to\eta_0.
\end{cases}
\end{gather}

\subsection{Riemann-Hilbert problem for $E$}

The error vector $E$ is defined in \eqref{error vector}, where the global parametrix, instead of \eqref{global parametrix}, at this stage is given by
\begin{gather}\label{global-parametrix-1}
P=
\begin{cases}
P^\infty\left(\lambda\right), &\mathrm{for}\,\,\,\lambda\in\mathbb{C}\setminus\overline{B\left(\pm\eta_2, \pm\eta_0, \pm\alpha\right)},  \v\\
P^{\eta_2}\left(\lambda\right), & \mathrm{for}\,\,\,\lambda\in B\left(\eta_2\right),  \v\\
P^{\eta_0}\left(\lambda\right), & \mathrm{for}\,\,\,\lambda\in B\left(\eta_0\right),  \v\\
P^{\alpha}\left(\lambda\right), & \mathrm{for}\,\,\,\lambda\in B\left(\alpha\right),  \v\\
\sigma_1P^{\eta_2}\left(-\lambda\right)\sigma_1, & \mathrm{for}\,\,\,\lambda\in B\left(-\eta_2\right), \v \\
\sigma_1P^{\eta_0}\left(-\lambda\right)\sigma_1, & \mathrm{for}\,\,\,\lambda\in B\left(-\eta_0\right), \v \\
\sigma_1P^{\alpha}\left(-\lambda\right)\sigma_1, & \mathrm{for}\,\,\,\lambda\in B\left(-\alpha\right),
\end{cases}
\end{gather}
with $B\left(\pm\eta_2, \pm\eta_0, \pm\alpha\right)=B(\eta_2)\cup B(-\eta_2)\cup B(\eta_0)\cup B(-\eta_0)\cup B(\alpha)\cup B(-\alpha)$.
\begin{figure}[!t]
\centering
\includegraphics[scale=0.35]{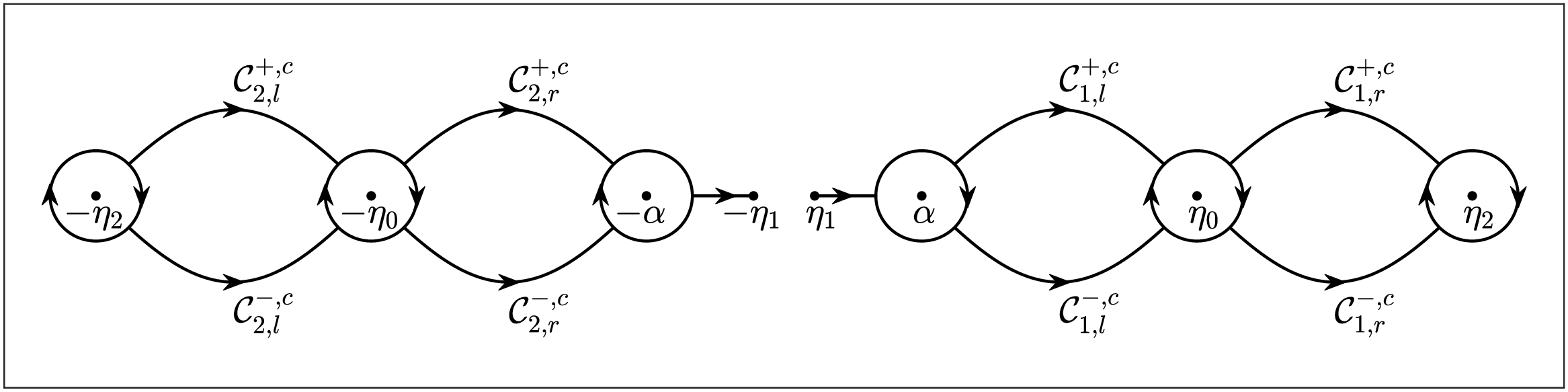}
\caption{Jump contours of the error vector $E$ in the region $\xi_{\mathrm{crit}}<\xi<\xi_0$.}
\label{ErrorE}
\end{figure}

The error vector $E$ solves the Riemann-Hilbert problem with properties as depicted in Figure \ref{ErrorE}.
$E$ is analytic in $\lambda$ for $\lambda\in\mathbb{C}\setminus\partial B\left(\pm\eta_2, \pm\eta_0, \pm\alpha\right)\cup\overline{\mathcal{C}_1^c}\cup\overline{\mathcal{C}_2^c}$, with $\mathcal{C}_1^c=\mathcal{C}_1\setminus \overline{B\left(\eta_2, \eta_0, \alpha\right)}$, $\mathcal{C}_2^c=\mathcal{C}_1\setminus \overline{B\left(-\eta_2, -\eta_0, -\alpha\right)}$, where $B\left(\eta_2, \eta_0, \alpha\right)=B\left(\eta_2\right)\cup B\left(\eta_0\right)\cup B\left(\alpha\right)$,  $B\left(-\eta_2, -\eta_0, -\alpha\right)=B\left(-\eta_2\right)\cup B\left(-\eta_0\right)\cup B\left(-\alpha\right)$ and
normalizes to
$
\begin{pmatrix}
1 &1
\end{pmatrix}
$
as $\lambda\to\infty$.
For $\lambda\in \partial B\left(\pm\eta_2, \pm\eta_0, \pm\alpha\right)\cup\mathcal{C}_1^c\cup\mathcal{C}_2^c$, $E$ admits continuous boundary values, denoted by $E_+$ and $E_-$, respectively. Let
\begin{gather}
\begin{aligned}
V^E_1&=P^\infty_-f_-^{\sigma_3}\mathrm{e}^{tp_-\sigma_3}\left(\mathcal{L}\left[-\mathrm{i}r\right]
-\mathbb{I}_2\right)\mathrm{e}^{-tp_+\sigma_3}f_+^{-\sigma_3}\left(P^\infty_+\right)^{-1},\\[0.5em]
V^E_2&=P^\infty_-f_-^{\sigma_3}\mathrm{e}^{tp_-\sigma_3}\left(\mathcal{U}\left[\mathrm{i}r\right]-\mathbb{I}_2\right)\mathrm{e}^{-tp_+\sigma_3}f_+^{-\sigma_3}\left(P^\infty_+\right)^{-1}.
\end{aligned}
\end{gather}
Then $E_+$ and $E_-$ have the following jump conditions
\begin{gather}\label{Error-Jump}
E_+=E_-
\begin{cases}
P^\infty \,\mathcal{U}^{tp}_f\left[ir^{-1}\right] \left(P^\infty\right)^{-1}, &\mathrm{for} \,\,\, \lambda\in\mathcal{C}^c_1,  \v\\
P^\infty \,\mathcal{L}^{tp}_f\left[-ir^{-1}\right] \left(P^\infty\right)^{-1}, &\mathrm{for} \,\,\, \lambda\in\mathcal{C}^c_2,  \v\\
P^{\eta_2}\left(P^\infty\right)^{-1}, &\mathrm{for} \,\,\, \lambda\in\partial B\left(\eta_2\right),  \v\\
P^{\eta_0}\left(P^\infty\right)^{-1}, &\mathrm{for} \,\,\, \lambda\in\partial B\left(\eta_0\right),  \v\\
P^{\alpha}\left(P^\infty\right)^{-1}, &\mathrm{for} \,\,\, \lambda\in\partial B\left(\alpha\right),   \v\\
\sigma_1P^{\eta_2}\left(-\lambda\right)\left(P^\infty\left(-\lambda\right)\right)^{-1}\sigma_1, &\mathrm{for} \,\,\, \lambda\in\partial B\left(-\eta_2\right), \v\\
\sigma_1P^{\eta_0}\left(-\lambda\right)\left(P^\infty\left(-\lambda\right)\right)^{-1}\sigma_1, &\mathrm{for} \,\,\, \lambda\in\partial B\left(-\eta_0\right),\v\\
\sigma_1P^{\alpha}\left(-\lambda\right)\left(P^\infty\left(-\lambda\right)\right)^{-1}\sigma_1, &\mathrm{for} \,\,\, \lambda\in\partial B\left(-\alpha\right),\v \\
\mathbb{I}_2+V^E_1, &\mathrm{for} \,\,\, \lambda\in \left(\eta_1, \alpha\right)\setminus B\left(\alpha \right), \v \\
\mathbb{I}_2+V^E_2, &\mathrm{for} \,\,\, \lambda\in \left(-\alpha, -\eta_1\right)\setminus B\left(-\alpha \right),
\end{cases}
\end{gather}

Near each self-intersection point, $E$ exhibits the following local behavior:
$
E=\mathcal{O}
\begin{pmatrix}
1 & 1
\end{pmatrix}
$
as $\lambda$ tends to each self-intersection  point.
Note that $P^\infty$ is formulated in \eqref{outer parametrix}.
$P^{\eta_2}$ is also formulated  in \eqref{parametrix-eta_2} and $P^\alpha$ in \eqref{parametrix-alpha-1} and \eqref{parametrix-alpha-2}. Thus,  both \eqref{small-norm-eta_2} and \eqref{small-norm-alpha} also hold. $P^{\eta_0}$ is established in \eqref{parametrix-eta_0-1}, \eqref{parametrix-eta_0-2}, \eqref{parametrix-eta_0-3}, and \eqref{parametrix-eta_0-4} for $r=r_0$ and in \eqref{parametrix-eta_0-a} and \eqref{parametrix-eta_0-b} for $r=r_c$. This leads to
\begin{gather}\label{small-norm-eta_0}
P^{\eta_0}\left(P^\infty\right)^{-1}=\mathbb{I}_2+\mathcal{O}\left(\frac{1}{t}\right), \quad \mathrm{as} \,\,\, t\to +\infty,
\end{gather}
which holds uniformly  for $\lambda\in\partial B\left(\eta_2\right)$. Again, the standard small norm argument is employed to derive  the asymptotic \eqref{asymptotic-middle} in the region $\xi\in\left(\xi_\mathrm{crit}, \xi_0\right)$.

\section{Long-time asymptotics: the region $\xi<\xi_{\mathrm{crit}}$}

For the region $\xi<\xi_{\mathrm{crit}}$, the Riemann-Hilbert problem for $Y$ is the same as that for the region $\xi_0<\xi<\eta_2^2$.
Thus, this section proceeds by introducing $g$- and $f$-functions for the conjugation. They are a slightly different from those introduced in the region  $\xi_0<\xi<\eta_2^2$.

\subsection{Riemann-Hilbert problems for $T$ and $S$}

The scalar $g$-function is analytic in $\lambda$ for $\lambda\in\mathbb{C}\setminus\left[-\eta_2, \eta_2\right]$ and normalizes to $\mathcal{O}\left(\lambda^{-1}\right)$ as $\lambda\to\infty$.
Its continuous boundary values $g_\pm$ are related by: $g_++g_- =2\theta$ for $\lambda\in\left(-\eta_2, -\eta_1\right)\cup\left(\eta_1, \eta_2\right)$;  $g_+-g_-=\Omega$  with  $\Omega=2\mathrm{i}\pi\eta_2\left(\eta_2^2+\eta_1^2-2\xi\right)\left/\right. eK\left(\eta_1\left/\right.\eta_2\right)$ for $\lambda\in\left(-\eta_1, \eta_1\right)$.
It follows that the $g$-function is expressed as \eqref{g-function},  where
 $R(s)$ is a branch of the complex function $\sqrt{\left(s^2-\eta_1^2\right)\left(s^2-\eta_2^2\right)}$ such that it is real and positive on $\left(\eta_2, +\infty\right)$ with branch cuts on the contours $\left(\eta_1, \eta_2\right)$ and $\left(-\eta_2, -\eta_1\right)$ and
$Q(s)=12s^4-(4\xi+6\eta_1^2+6\eta_2^2)s^2+2\eta_2^2(\eta_1^2-\eta_2^2+2\xi+(\eta_1^2+\eta_2^2-2\xi)eE(\eta_1/\eta_2)\left/\right.eK(\eta_1/\eta_2))$.
The $f$-function is analytic in $\lambda$ for $\lambda\in\mathbb{C}\setminus\left[-\eta_2, \eta_2\right]$ and normalizes to 1 as $\lambda\to\infty$.
Its continuous boundary values $g_\pm$ are related by: $f_+f_- =r^{-1}$ for $\lambda\in\left(\eta_1, \eta_0\right)\cup\left(\eta_0, \eta_2\right)$;
$f_+f_- =r$ for $\lambda\in\left(-\eta_2, -\eta_0\right)\cup\left(-\eta_0, -\eta_1\right)$; $f_+f_-^{-1}=\mathrm{e}^\Delta$  with  $\Delta=\int_{\eta_1}^{\eta_2}\left(\log r(s)\right)/R_+\left(s\right)\,\mathrm{d}s\left/\right.\int_0^{\eta_1} 1\left/\right.R(s)\mathrm{d}s\in \mathrm{i}\mathbb{R}$ for $\lambda\in\left(-\eta_1, \eta_1\right)$.
Divided by $R_+$, the $f$-function is obtained by taking the logarithm and using Plemelj's formula as follows
\begin{gather}
f=\exp\left\{\frac{R}{2\pi \mathrm{i}}\left(\int_{\eta_1}^{\eta_2}-\frac{\log r(s)}{R_+\left(s\right)}\frac{\mathrm{d}s}{s-\lambda}+\int_{-\eta_2}^{-\eta_1}\frac{\log r(s)}{R_+\left(s\right)}\frac{\mathrm{d}s}{s-\lambda}+\int_{-\eta_1}^{\eta_1}\frac{\Delta}{R(s)}\frac{\mathrm{d}s}{s-\lambda}\right)\right\}.
\end{gather}
With the definitions of the $g$- and $f$-functions, the conjugation is performed as in \eqref{Conjugation}, which determines a $1\times 2$ vector-valued function $T$, solving a Riemann-Hilbert problem. 	
$T$ is analytic in $\lambda$ for $\lambda\in\mathbb{C}\setminus\left(\left[-\eta_2, \eta_2\right]\right)$ and normalizes to $
\begin{pmatrix}
1& 1
\end{pmatrix}$
as $\lambda\to\infty$.
For  $\lambda\in\left(-\eta_2, \eta_2\right)\setminus\left\{\pm\eta_1, \pm\eta_0\right\}$, $T$ admits continuous boundary values, which are related by the following jump conditions
\begin{gather}
T_+=T_-
\begin{cases}
\mathcal{U}^{tp_-}_{f_-}\left[\mathrm{i}r^{-1}\right]\left(-\mathrm{i}\sigma_1\right)\mathcal{U}^{tp_+}_{f_+}\left[\mathrm{i}r^{-1}\right], &\mathrm{for}\,\,\, \lambda\in\left(\eta_1, \eta_0\right)\cup\left(\eta_0, \eta_2\right),\\[0.5em]
\mathcal{L}^{tp_-}_{f_-}\left[-\mathrm{i}r^{-1}\right]\left(\mathrm{i}\sigma_1\right)\mathcal{L}^{tp_+}_{f_+}\left[-\mathrm{i}r^{-1}\right], &\mathrm{for}\,\,\, \lambda\in\left(-\eta_2, -\eta_0\right)\cup\left(-\eta_0, -\eta_1\right),\\[0.5em]
\mathrm{e}^{\left(\Omega t+\Delta\right)\sigma_3}, &\mathrm{for}\,\,\, \lambda\in\left(-\eta_1, \eta_1\right).
\end{cases}
\end{gather}
The local behaviors of $T$ near $\pm\eta_2$ and $\pm\eta_0$ are the same as those in the region of $\xi_{\mathrm{crit}}<\xi<\xi_0$. $T$ exhibits the following local behavior as $\lambda\to\pm\eta_1$
\begin{gather}
T=
\begin{cases}
\mathcal{O}
\begin{pmatrix}
 \left|\lambda\mp\eta_1\right|^{\mp\beta_1/2}&\left|\lambda\mp\eta_1\right|^{\pm\beta_1/2}
\end{pmatrix}
, & \mathrm{if} \,\,\,\beta_1\in\left(0, +\infty\right), \\[0.5em]
\mathcal{O}
\begin{pmatrix}
 \left|\lambda\mp\eta_1\right|^{\beta_1/2}&\left|\lambda\mp\eta_1\right|^{\beta_1/2}
\end{pmatrix}
, & \mathrm{if} \,\,\,\beta_1\in\left(-1, 0\right).
\end{cases}
\end{gather}

\begin{figure}[!t]
\centering
\includegraphics[scale=0.36]{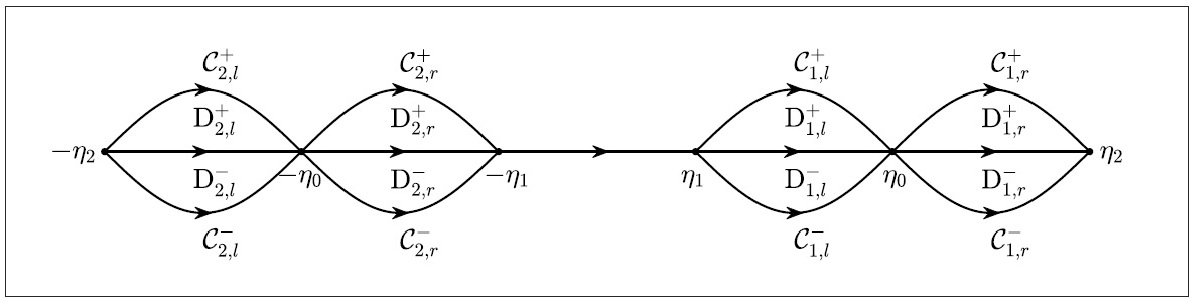}
\caption{Contour deformation by opening lenses in the region $\xi<\xi_{\mathrm{crit}}$.}
\label{Deformation4}
\end{figure}

The contour deformation of the Riemann-Hilbert problem by opening lenses is depicted in Figure \ref{Deformation4}.
The domains $\mathrm{D}^+_{1, l}$ and $\mathrm{D}^-_{1, l}$ are lenses, respectively, above and below $(\eta_1, \eta_0)$.
The domains $\mathrm{D}^+_{1, r}$ and $\mathrm{D}^-_{1, r}$ are lenses, respectively, above and below $(\eta_0, \eta_2)$.
The domains $\mathrm{D}^+_{2, l}$ and $\mathrm{D}^-_{2, l}$ are lenses, respectively, above and below $(-\eta_2, -\eta_0)$.
The domains $\mathrm{D}^+_{2, r}$ and $\mathrm{D}^-_{2, r}$ are lenses, respectively, above and below $(-\eta_0, -\eta_1)$.
For convenience, define
$\mathrm{D}_{1, l}=\mathrm{D}_{1, l}^+\cup \mathrm{D}_{1, l}^-$,
$\mathrm{D}_{1, r}=\mathrm{D}_{1, r}^+\cup \mathrm{D}_{1, r}^-$,
$\mathrm{D}_1^+=\mathrm{D}_{1, l}^+\cup \mathrm{D}_{1, r}^+$,
$\mathrm{D}_1^-=\mathrm{D}_{1, l}^-\cup \mathrm{D}_{1, r}^-$,
$\mathrm{D}_{2, l}=\mathrm{D}_{2, l}^+\cup \mathrm{D}_{2, l}^-$,
$\mathrm{D}_{2, r}=\mathrm{D}_{2, r}^+\cup \mathrm{D}_{2, r}^-$,
$\mathrm{D}_2^+=\mathrm{D}_{2, l}^+\cup \mathrm{D}_{2, r}^+$,
$\mathrm{D}_2^-=\mathrm{D}_{2, l}^-\cup \mathrm{D}_{2, r}^-$,
$\mathrm{D}_1=\mathrm{D}_{1, l}\cup \mathrm{D}_{1, r}$,
and
$\mathrm{D}_2=\mathrm{D}_{2, l}\cup \mathrm{D}_{2, r}$.
Denote $\mathrm{D}_\mathrm{o}$ as the domain outside these lenses, i.e.,
$\mathrm{D}_\mathrm{o}=\mathbb{C}\setminus \overline{\mathrm{D}_1\cup \mathrm{D}_2\cup(-\eta_1, \eta_1)}$
and $\mathrm{D}_\mathrm{o}=\mathrm{D}_\mathrm{o}^+\cup \mathrm{D}_\mathrm{o}^- \cup (\eta_2, +\infty)\cup(-\infty, -\eta_2)$,
where $\mathrm{D}_\mathrm{o}^+$ and $\mathrm{D}_\mathrm{o}^-$ denote the parts in the upper half-plane and lower half-plane, respectively.
The $1\times 2$ vector-valued function $S$ is defined as follows:
$S=T\mathcal{U}^{tp}_{f}\left[\mathrm{i}r^{-1}\right]^{-1}$ for $\lambda\in\mathrm{D}_1^+$;
$S=T\mathcal{U}^{tp}_{f}\left[\mathrm{i}r^{-1}\right]$ for $\lambda\in\mathrm{D}_1^-$;
$S=T\mathcal{L}^{tp}_{f}\left[-\mathrm{i}r^{-1}\right]^{-1}$ for $\lambda\in\mathrm{D}_2^+$;
$S=T\mathcal{L}^{tp}_{f}\left[-\mathrm{i}r^{-1}\right]$ for $\lambda\in\mathrm{D}_2^-$;
$S=T$ for $\lambda\in\mathrm{D}_\mathrm{o}$,
where $p$ is still defined in \eqref{definition-p}.
The defining function $S$ solves the following Riemann-Hilbert problem.
$S$ is analytic in $\lambda$ for $\lambda\in\mathbb{C}\setminus\left(\left[-\eta_2, \eta_2\right]\cup\overline{\mathcal{C}_1}\cup\overline{\mathcal{C}_2}\right)$, with $\mathcal{C}_1=\mathcal{C}_{1, l}^+\cup\mathcal{C}_{1, r}^+\cup\mathcal{C}_{1, l}^-\cup\mathcal{C}_{1, r}^-$ and $\mathcal{C}_2=\mathcal{C}_{2, l}^+\cup\mathcal{C}_{2, r}^+\cup\mathcal{C}_{2, l}^-\cup\mathcal{C}_{2, r}^-$,
and normalizes to
$
\begin{pmatrix}
1& 1
\end{pmatrix}
$
as $\lambda\to\infty$.
For  $\lambda\in\left(\eta_2, \eta_2\right)\cup\mathcal{C}_1\cup\mathcal{C}_2\setminus\left\{\pm\eta_1, \pm\eta_0\right\}$, $S$ admits continuous boundary values  denoted by $S_+$ and $S_-$, respectively. These values are related by the following jump conditions
\begin{gather}
S_+=S_-
\begin{cases}
\mathcal{U}^{tp}_{f}\left[\mathrm{i}r^{-1}\right], &\mathrm{for}\,\,\, \lambda\in\mathcal{C}_1,\\[0.5em]
\mathcal{L}^{tp}_{f}\left[-\mathrm{i}r^{-1}\right], &\mathrm{for}\,\,\, \lambda\in\mathcal{C}_2,\\[0.5em]
-\mathrm{i}\sigma_1, &\mathrm{for}\,\,\, \lambda\in\left(\eta_1, \eta_0\right)\cup\left(\eta_0, \eta_2\right),\\[0.5em]
\mathrm{i}\sigma_1, &\mathrm{for}\,\,\, \lambda\in\left(-\eta_2, -\eta_0\right)\cup\left(-\eta_0, -\eta_1\right),\\[0.5em]
\mathrm{e}^{\left(\Omega t+\Delta\right)\sigma_3}, &\mathrm{for}\,\,\, \lambda\in\left(-\eta_1, \eta_1\right).
\end{cases}
\end{gather}
The local behaviors of $S$ near $\pm\eta_2$ and $\pm\eta_0$ are the same as those in the case of $\xi_{\mathrm{crit}}<\xi<\xi_0$. $S$ exhibits the following local behavior as $\lambda\to\pm\eta_1$:
if $\beta_1>0$, the local behavior is expressed as
\begin{gather}
S=
\begin{cases}
\mathcal{O}
\begin{pmatrix}
 \left|\lambda\mp\eta_1\right|^{\mp\beta_1/2}&\left|\lambda\mp\eta_1\right|^{\pm\beta_1/2}
\end{pmatrix}
, &\mathrm{as} \,\,\,\lambda\in\mathrm{D}_{\mathrm{o}}\to\pm\eta_1,\\[0.5em]
\mathcal{O}
\begin{pmatrix}
 \left|\lambda\mp\eta_1\right|^{-\beta_1/2}&\left|\lambda\mp\eta_1\right|^{-\beta_1/2}
\end{pmatrix}
, &\mathrm{as} \,\,\,\lambda\in\mathrm{D}_1\cup \mathrm{D}_2\to\pm\eta_1,
\end{cases}
\end{gather}
and if $\beta_1\in\left(-1, 0\right)$, the local behavior is formulated as
\begin{gather}
S=
\mathcal{O}
\begin{pmatrix}
 \left|\lambda\mp\eta_1\right|^{\beta_1/2}&\left|\lambda\mp\eta_1\right|^{\beta_1/2}
\end{pmatrix}, \quad
\mathrm{as} \,\,\,\lambda\in\mathrm{D}_1\cup\mathrm{D}_2\cup \mathrm{D}_{\mathrm{o}}\to\pm\eta_1.
\end{gather}

\subsection{Local parametrix and Riemann-Hilbert problem for $E$}

The error vector $E$ is defined in \eqref{error vector}, where the global parametrix  is the same as that written in \eqref{global-parametrix-1} with $\alpha$ replaced by $\eta_1$ in this case,  instead of $\alpha$.
The error vector $E$ solves a Riemann-Hilbert problem with properties similar to the region  $\xi\in\left(\xi_\mathrm{crit}, \xi_0\right)$ except for the endpoint $\eta_1$. $E$ is analytic in $\lambda$ for $\lambda\in\mathbb{C}\setminus\partial B\left(\pm\eta_2, \pm\eta_0, \pm\eta_1\right)\cup\overline{\mathcal{C}_1^c}\cup\overline{\mathcal{C}_2^c}$, with $\mathcal{C}_1^c=\mathcal{C}_1\setminus \overline{B\left(\eta_2, \eta_0, \eta_1\right)}$, $\mathcal{C}_2^c=\mathcal{C}_1\setminus \overline{B\left(-\eta_2, -\eta_0, -\eta_1\right)}$, $B\left(\eta_2, \eta_0, \eta_1\right)=B\left(\eta_2\right)\cup B\left(\eta_0\right)\cup B\left(\eta_1\right)$,  and $B\left(-\eta_2, -\eta_0, -\eta_1\right)=B\left(-\eta_2\right)\cup B\left(-\eta_0\right)\cup B\left(-\eta_1\right)$.
The error vector $E$ normalizes to
$
\begin{pmatrix}
1 &1
\end{pmatrix}
$
as $\lambda\to\infty$.
For $\lambda\in \partial B\left(\pm\eta_2, \pm\eta_0, \pm\alpha\right)\cup\mathcal{C}_1^c\cup\mathcal{C}_2^c$, $E$ admits continuous boundary values, denoted by $E_+$ and $E_-$, respectively. These values are related by \eqref{Error-Jump} other than at $\lambda\in\partial B\left(\pm\eta_1\right)$, as depicted in Figure \ref{ErrorE1}
\begin{gather}\label{ErrorE2}
E_+=E_-
\begin{cases}
P^{\eta_1}\left(P^\infty\right)^{-1}, &\mathrm{for} \,\,\, \lambda\in\partial B\left(\eta_1\right),  \v \\
\sigma_1P^{\eta_1}\left(-\lambda\right)\left(P^\infty\left(-\lambda\right)\right)^{-1}\sigma_1, &\mathrm{for} \,\,\, \lambda\in\partial B\left(-\eta_1\right).
\end{cases}
\end{gather}
Near each self-intersection point, $E$ exhibits the following local behavior:
$
E=\mathcal{O}
\begin{pmatrix}
1 & 1
\end{pmatrix}
$
as $\lambda$ tends to each self-intersection  point.

In \eqref{ErrorE2}, $P^{\eta_1}$ is the local parametrix in the neighbourhood of $\lambda=\eta_1$, which is formulated as follows:
for $\lambda\in\left(\mathrm{D}_{1, l}^+\cup \mathrm{D}_\mathrm{o}^+\right)\cap B(\eta_1)$, the local parametrix, denoted as $P^{\eta_1}$, is
\begin{gather}\label{parametrix-eta_1-1}
P^{\eta_1}=P^\infty\left(\frac{\mathrm{e}^{\pi\mathrm{i}/4-\Omega t/2}}{fd}\right)^{\sigma_3}\sigma_2C\zeta_{\eta_1}^{-\sigma_3/4}M^{\mathrm{mB}}(\zeta_{\eta_1}, \beta_1)\mathrm{e}^{-\sqrt{\zeta_{\eta_1}}\sigma_3}\sigma_2\left(\frac{\mathrm{e}^{\pi\mathrm{i}/4-\Omega t/2}}{fd}\right)^{-\sigma_3},
\end{gather}
with $\zeta_{\eta_2}=t^2\left(p+\Omega/2\right)^2$;
for $\lambda\in\left(\mathrm{D}_{1, l}^-\cup \mathrm{D}_\mathrm{o}^-\right) \cap B(\eta_1)$, the local parametrix $P^{\eta_1}$ is
\begin{gather}\label{parametrix-eta_1-2}
P^{\eta_1}=P^\infty\left(\frac{\mathrm{e}^{\pi\mathrm{i}/4+\Omega t/2}}{fd}\right)^{\sigma_3}\sigma_2C\zeta_{\eta_1}^{-\sigma_3/4}M^{\mathrm{mB}}(\zeta_{\eta_1}, \beta_1)\mathrm{e}^{-\sqrt{\zeta_{\eta_1}}\sigma_3}\sigma_2\left(\frac{\mathrm{e}^{\pi\mathrm{i}/4+\Omega t/2}}{fd}\right)^{-\sigma_3},
\end{gather}
with $\zeta_{\eta_2}=t^2\left(p-\Omega/2\right)^2$.
In expressions \eqref{parametrix-eta_1-1} and \eqref{parametrix-eta_1-2}, $M^{\mathrm{mB}}(\zeta_{\eta_1}, \beta_1)$ is formulated as shown in \eqref{modified-Bessel-1}, \eqref{modified-Bessel-2}, and \eqref{modified-Bessel-3}, where $\zeta_{\eta_2}$ is replaced by $\zeta_{\eta_1}$ and $\beta_2$ by $\beta_1$.
The function $d$  is expressed as follows:
for the first type of generalized reflection coefficient, $r=r_0$, $d=(\eta_1-\lambda)^{\beta_1/2}(\eta_2-\lambda)^{\beta_2/2}\left|\lambda-\eta_0\right|^{\beta_0/2}\gamma(\lambda)^{1/2}$;
for the second type of generalized reflection coefficient, $r=r_c$, $d=(\eta_1-\lambda)^{\beta_1/2}(\eta_2-\lambda)^{\beta_2/2}\chi_c(\lambda)^{1/2}\gamma(\lambda)^{1/2}$.

\begin{figure}[!t]
\centering
\includegraphics[scale=0.36]{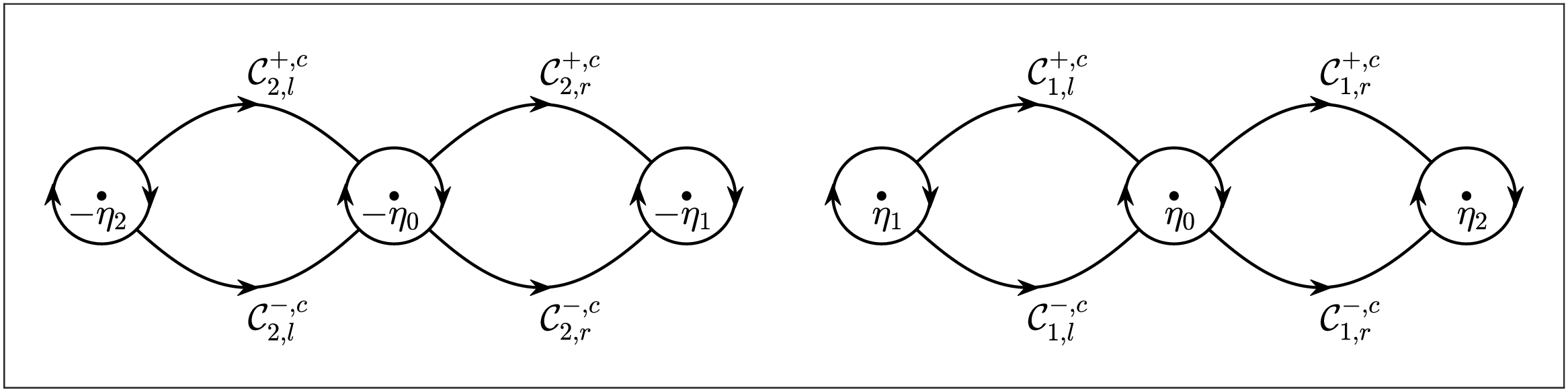}
\caption{Jump contours of the error vector $E$ in the region $\xi<\xi_{\mathrm{crit}}$.}
\label{ErrorE1}
\end{figure}

In the neighborhood of $\lambda=\eta_1$, the matrix $P^{\eta_1}$ satisfies a Riemann-Hilbert problem matching well $S$.  $P^{\eta_1}$ is analytic in $\lambda$ for $\lambda\in\left(\mathrm{D}_1\cup\mathrm{D}_\mathrm{o}\right)\cap B(\eta_1)$, and as $t\to\infty$, it normalizes to
\begin{gather}
P^{\eta_1}\left(P^\infty\right)^{-1}=\mathbb{I}_2+\mathcal{O}\left(\frac{1}{t}\right),
\end{gather}
which holds uniformly  for $\lambda\in\partial B\left(\eta_1\right)$.
For $\lambda\in\left(\mathcal{C}_1\cup \left(0, \eta_0\right)\right)\cap B(\eta_1)$, the matrix $P^{\eta_1}$ admits continuous boundary values, which are related by the following jump relation
\begin{gather}
P^{\eta_1}_+=P^{\eta_1}_-
\begin{cases}
\mathcal{U}^{tp}_{f}\left[\mathrm{i}r^{-1}\right], &\mathrm{for}\,\,\, \lambda\in\mathcal{C}_1\cap B(\eta_1),\\[0.5em]
-\mathrm{i}\sigma_1, &\mathrm{for}\,\,\, \lambda\in\left(\eta_1, \eta_0\right)\cap B(\eta_1),\\[0.5em]
\mathrm{e}^{\left(\Omega t+\Delta\right)\sigma_3}, &\mathrm{for}\,\,\, \lambda\in\left(0, \eta_1\right)\cap B(\eta_1).
\end{cases}
\end{gather}
In the neighborhood of $\lambda=\eta_1$, $S$ exhibits the following local behavior:
if $\beta_1>0$, the local behavior is expressed as
\begin{gather}
P^{\eta_1}=
\begin{cases}
\mathcal{O}
\begin{pmatrix}
 \left|\lambda-\eta_1\right|^{-\beta_1/2}&\left|\lambda-\eta_1\right|^{\beta_1/2}\\[0.5em]
  \left|\lambda-\eta_1\right|^{-\beta_1/2}&\left|\lambda-\eta_1\right|^{\beta_1/2}
\end{pmatrix}
, &\mathrm{as} \,\,\,\lambda\in\mathrm{D}_{\mathrm{o}}\to\eta_1,\\[2em]
\mathcal{O}
\begin{pmatrix}
 \left|\lambda-\eta_1\right|^{-\beta_1/2}&\left|\lambda-\eta_1\right|^{-\beta_1/2}\\[0.5em]
 \left|\lambda-\eta_1\right|^{-\beta_1/2}&\left|\lambda-\eta_1\right|^{-\beta_1/2}
\end{pmatrix}
, &\mathrm{as} \,\,\,\lambda\in\mathrm{D}_1\to\eta_1,
\end{cases}
\end{gather}
and if $\beta_1\in\left(-1, 0\right)$, the local behavior is formulated as
\begin{gather}
P^{\eta_1}=
\mathcal{O}
\begin{pmatrix}
 \left|\lambda-\eta_1\right|^{\beta_1/2}&\left|\lambda-\eta_1\right|^{\beta_1/2} \\[0.5em]
  \left|\lambda-\eta_1\right|^{\beta_1/2}&\left|\lambda-\eta_1\right|^{\beta_1/2}
\end{pmatrix}, \quad
\mathrm{as} \,\,\,\lambda\in\mathrm{D}_1\cup \mathrm{D}_{\mathrm{o}}\to\eta_1.
\end{gather}

Note that $P^\infty$ is formulated in \eqref{outer parametrix}, with $\alpha$ replaced by $\eta_1$. $P^{\eta_2}$ is similarly formulated in \eqref{parametrix-eta_2}. For the first type of generalized reflection coefficient, $r = r_0$, $P^{\eta_0}$ is established in \eqref{parametrix-eta_0-1}, \eqref{parametrix-eta_0-2}, \eqref{parametrix-eta_0-3}, and \eqref{parametrix-eta_0-4}; for the second type, $r = r_c$, it is given in \eqref{parametrix-eta_0-a} and \eqref{parametrix-eta_0-b}. Consequently, both \eqref{small-norm-eta_2} and \eqref{small-norm-eta_0} hold. Finally, the asymptotic behavior \eqref{asymptotic-left} of the soliton gas in the region $\xi \in (-\infty, \xi_{\mathrm{crit}})$ follows from performing the standard small norm argument.

\section{Conclusions and discussions}

In summary, we have investigated the long-time asymptotic behaviors of KdV soliton gases, characterized by the generalized reflection coefficients in the context of the Riemann-Hilbert problems.
As an extension, two types of generalized reflection coefficients are considered: the first one, $r_0=(\lambda-\eta_1)^{\beta_1}(\eta_2-\lambda)^{\beta_2}\left|\lambda-\eta_0\right|^{\beta_0}\gamma(\lambda)$, and the second one, $r_c=(\lambda-\eta_1)^{\beta_1}(\eta_2-\lambda)^{\beta_2}\chi_c\left(\lambda, \eta_0\right)\gamma\left(\lambda\right)$.
The primary challenge lies in constructing the local parametrices near the endpoints $\eta_1$ and $\eta_2$, as well as the newly introduced singularity $\eta_0$.
For the endpoints $\eta_j$ with $j=1, 2$, the modified Bessel functions of the first and second kind, with index $\beta_j$, are employed to construct the corresponding local parametrix $P^{\eta_j}$.
For the first type of generalized reflection coefficient, $r_0$, the local parametrix $P^{\eta_0}$ is constructed using modified Bessel functions with index $(\beta_0\pm 1)/2$.
For the second type of generalized reflection coefficient, $r_c$, the local parametrix $P^{\eta_0}$ is constructed using confluent hypergeometric functions. Moreover, these results can also be extended to the case that these reflection coefficients accommodate an arbitrary number $n$ of singularities $\{\eta_{0,j}\}_1^n$, given by Eqs.~\eqref{general-0} and \eqref{general-1}.

Several challenges remain. For example, i) As noted in \cite{15}, deriving the rigorous asymptotics of the soliton gas in the presence of two nontrivial reflection coefficients remains a challenging problem; ii) Addressing the limit procedure when the discrete spectra accumulate in disconnected components of the imaginary axis poses another significant challenge; iii)
The sine-Gordon equation and the Camassa-Holm equation are two important and fundamental integrable models that admit multi-kink and multi-peakon solutions, respectively.
Studying the asymptotics of the corresponding kink gases and peakon gases using the Deift-Zhou method is still an interesting avenue of research.
For more important open problems, one can refer to the review paper \cite{69}, which summarizes various fundamental questions inspired by the exciting theoretical and experimental challenges.



\addcontentsline{toc}{section}{Acknowledgments}

\vspace{0.2in}
\noindent {\bf Acknowledgments}

\vspace{0.05in}
We thank K. D. T.-R. McLaughlin very much for his valuable suggestions and comments about the original manuscript.
This work of G.Z. was supported by the National Natural Science Foundation of China (Grant No. 12201615).
The work of Z.Y. was supported by the National Natural Science Foundation of China (No. 11925108).










\end{document}